\documentclass[twocolumn]{aastex631}
\usepackage[mathlines]{lineno}
\usepackage{amsmath}
\usepackage{booktabs}
\usepackage{microtype}
\usepackage{silence}
\usepackage{natbib}
\usepackage{graphicx}
\usepackage[caption=false]{subfig}
\usepackage[noabbrev,capitalize]{cleveref}
\usepackage{glossaries-extra}
\glsdisablehyper
  \usepackage{siunitx}
\usepackage{placeins}
\usepackage{xurl}
\usepackage[gobble=auto]{pythontex}
\setpythontexcontext{textwidth=\the\textwidth, linewidth=\the\linewidth, columnwidth=\the\columnwidth}
\begin{pycode}
    sys.path.append('/Users/golmschenk/Code/lyla')
    sys.path.append('/Users/golmschenk/Code/ramjet')
    from lyla.pythontex_liaison import PythontexLiaison
    from lyla.figure import create_latex_figure_from_bokeh_figure, create_latex_figure_from_bokeh_layout
    PythontexLiaison.pytex = pytex
\end{pycode}

\usepackage{etoolbox}
\makeatletter
\patchcmd\H@refstepcounter{\protected@edef}{\protected@xdef}{}{}
\makeatother
\makeatletter
\patchcmd\linenumberpar{\@LN@parpgbrk}{\penalty\@LN@parpgpen\relax}{}{}
\makeatother

\newcommand{\eg}{\textit{e.g.}}
\newcommand{\ie}{\textit{i.e.}}
\newcommand{\approximately}{\raisebox{0.5ex}{\texttildelow}}

 \setabbreviationstyle[acronym]{long-short-user}

\newacronym[user1={\citealp{krizhevsky2012imagenet}}]{CNN}{CNN}{convolutional neural network}
\newacronym[user1={\eg, \citealp{lecun2015deep}}]{DNN}{DNN}{deep neural network}
\newacronym{ML}{ML}{machine learning}
\newacronym{NN}{NN}{neural network}
\newacronym[user1={\citealp{glorot2011deep}}]{ReLU}{ReLU}{rectified linear unit}

\newacronym{DS}{$\delta$ Sct}{Delta Scuti}
\newacronym{FFI}{FFI}{full-frame image}
\newacronym[user1={\citealp{brown2018gaia}}]{Gaia}{Gaia}{the Gaia Mission}
\newacronym{HR}{HR}{Hertzsprung-Russell}
\newacronym{MS}{MS}{main sequence}
\newacronym{LC}{light curve}{light curve}
\glsunset{LC}
\newacronym{RLOF}{RLOF}{Roche lobe overflow}
\newacronym{sdB}{sdB}{blue subdwarf}
\newacronym[user1={\citealp{ricker2014transiting}}]{TESS}{\textit{TESS}}{the \textit{Transiting Exoplanet Survey Satellite}}
\newacronym[user1={\citealp{stassun2018tess}}]{TIC}{TIC}{\textit{TESS} input catalog}
\newacronym{WD}{WD}{white dwarf}
\newacronym{ZAMS}{ZAMS}{zero-age main sequence}
\newacronym{TPR}{TPR}{true positive rate}
\newacronym{FPR}{FPR}{false positive rate}
 
\begin{document}
    
\title{Short-Period Variables in TESS Full-Frame Image Light Curves\\ Identified via Convolutional Neural Networks}

\newcommand{\gsfcAffiliationString}{NASA Goddard Space Flight Center, Greenbelt, MD 20771, USA}
\newcommand{\cuaAffiliationString}{Department of Physics, The Catholic University of America, Washington, DC 20064, USA}
\newcommand{\umdAffiliationString}{Department of Astronomy, University of Maryland, College Park, MD 20742, USA}
\newcommand{\orauAffiliationString}{Oak Ridge Associated Universities, Oak Ridge, TN 37830, USA} 
\author[0000-0001-8472-2219]{Greg Olmschenk}
\affiliation{\gsfcAffiliationString}
\affiliation{\umdAffiliationString}

\author{Richard K. Barry}
\affiliation{\gsfcAffiliationString}

\author[0000-0003-2267-1246]{Stela Ishitani Silva}
\affiliation{\gsfcAffiliationString}
\affiliation{\cuaAffiliationString}

\author[0000-0002-2942-8399]{Jeremy D. Schnittman}
\affiliation{\gsfcAffiliationString}

\author[0009-0006-9864-0517]{Agnieszka M. Cieplak}
\affiliation{Space Sciences Laboratory, 7 Gauss Way, University of California, Berkeley, CA 94720-7450, USA}

\author[0000-0003-0501-2636]{Brian P. Powell}
\affiliation{\gsfcAffiliationString}

\author[0000-0002-0493-1342]{Ethan Kruse}
\affiliation{\gsfcAffiliationString}

\author[0000-0001-7139-2724]{Thomas Barclay}
\affiliation{\gsfcAffiliationString}

\author{Siddhant Solanki}
\affiliation{\gsfcAffiliationString}

\author{Bianca Ortega}
\affiliation{\gsfcAffiliationString}

\author{John Baker}
\affiliation{\gsfcAffiliationString}

\author{Yesenia Helem Salinas Mamani}
\affiliation{Computer Science Department, School of Engineering, Pontificia Universidad Católica de Chile, Chile}

\received{2023-12-29}
\revised{2024-05-01}
\accepted{2024-06-10}
\published{2024-07-19}
\submitjournal{The Astronomical Journal}     \begin{pycode}
    from main_resources.generate_variables_file import generate_variables_file
    generate_variables_file()
    from results_resources.results_statistics_variables import generate_results_statistics_variables
    generate_results_statistics_variables()
    from results_resources.temperature_color_mapper import generate_high_temperature_latex_variables
    generate_high_temperature_latex_variables()
\end{pycode}
\newcommand{\numberOfShortPeriodVariables}{14156}
\newcommand{\numberOfTargetsInRedDwarfBinaryCluster}{6563}
\newcommand{\numberOfTargetsInDeltaScutiPrimaryRidgeCluster}{5637}
\newcommand{\numberOfTargetsInDeltaScutiSecondRidgeCluster}{1452}
\newcommand{\numberOfTargetsInBothDeltaScutiClusters}{7089}
\newcommand{\highTemperatureScaleLimit}{11000}
\newcommand{\highTemperatureColor}{red}
\newcommand{\exampleLightCurveTicIdZero}{149989733}
\newcommand{\exampleLightCurveSectorZero}{10}
\newcommand{\exampleLightCurvePeriodInHoursZero}{1.326}
\newcommand{\exampleLightCurveTicIdOne}{159971257}
\newcommand{\exampleLightCurveSectorOne}{23}
\newcommand{\exampleLightCurvePeriodInHoursOne}{3.338}
 \newcommand{\clusterColorRedDwarfBinary}{red}
\newcommand{\clusterColorDeltaScutiPrimaryRidge}{blue}
\newcommand{\clusterColorDeltaScutiSecondRidge}{yellow}
\newcommand{\clusterColorNone}{gray}
\newcommand{\implicitThreshold}{0.99796987}
\newcommand{\knownCount}{13647}
\newcommand{\knownAboveThreshold}{1363}
\newcommand{\randomTwoMinuteAboveThreshold}{16}
\newcommand{\twoMinuteFalsePositivePercent}{1.174\%}
 
\begin{abstract}
    The \textit{Transiting Exoplanet Survey Satellite} (\textit{TESS}) mission measured light from stars in \approximately85\% of the sky throughout its two-year primary mission, resulting in millions of \textit{TESS} 30-minute cadence light curves to analyze in the search for transiting exoplanets. To search this vast dataset, we aim to provide an approach that is computationally efficient, produces accurate predictions, and minimizes the required human search effort. We present a convolutional neural network that we train to identify short period variables. To make a prediction for a given light curve, our network requires no prior target parameters identified using other methods. Our network performs inference on a \textit{TESS} 30-minute cadence light curve in \approximately5ms on a single GPU, enabling large scale archival searches. We present a collection of \numberOfShortPeriodVariables{} short-period variables identified by our network. The majority of our identified variables fall into two prominent populations, one of close orbit main sequence binaries and another of Delta Scuti stars. Our neural network model and related code is additionally provided as open-source code for public use and extension.
\end{abstract}
     \section{Introduction}

The volume of astronomical photometric datasets is expanding rapidly. Due to their size, these datasets often contain data that no human eye has ever nor will ever see. Because of this, automated systems that can filter irrelevant information and identify interesting phenomena is imperative. In this work, we describe our development and application of a \gls{NN} to automatically identify short-period variables in \gls{TESS} \gls{FFI} data.

Modern wide-field surveys such as \gls{TESS}, the Optical Gravitational Lensing Experiment~\citep{sosz2015} and the Zwicky Transient Facility~\citep{ZTF2019} offer the possibility of developing and testing population synthesis models that provide insights into the distribution of variables in the Galaxy and they can help describe stellar evolutionary processes.

The primary goal of the \gls{TESS} mission, the observatory whose data was used for this work, is to detect the signature of planets as they transit in front of their host stars. Launched in April 2018, \gls{TESS} is performing a near all-sky photometric survey intended to identify planets with bright enough host stars to enable mass estimation from ground-based radial velocity measurements. Importantly, due to \gls{TESS}'s high observational cadence, these data are also well-suited to search for various other forms of short-duration, time-domain astrophysical phenomena.

The \gls{TESS} observatory is positioned in a high-Earth elliptical orbit of 13.7 days. \gls{TESS} completed its 2-year primary mission in July 2020. While a principle data product of \gls{TESS}'s primary mission is the 2-minute cadence photometry of more than 200,000 stars, of more relevance for this work is the \gls{FFI} data. During the primary mission, at a 30-minute cadence, \gls{TESS} took flux measurements of its entire field of view (\ang{24}\texttimes\ang{96}) resulting in the \gls{FFI} dataset. This \gls{FFI} data includes approximately 85\% of the sky and observes billions of sources~\citep{ricker2014transiting}. Searching this large-scale dataset requires a robust and computationally fast method and \glspl{NN} provide one such approach.

Largely due to their potential to approximate any given function~\citep{cybenko1989approximation,leshno1993multilayer,zhou2020universality}, in recent years, \glspl{DNN} have come to dominate the field of \gls{ML}. In the case of photometric data, a common function these methods may be tasked to learn is to predict physical classifications of the sources from observed flux measurements. Both \gls{ML} and non-\gls{ML} algorithms can only approximate the unknown ideal function that can perform this conversion perfectly.

When a \gls{NN} is trained to learn this function, it attempts to learn optimal data transformations~\citep{rumelhart1986learning}, which can potentially lead to more accurate classifications than manually designed methods, often because these manually designed methods frequently omit data during processing. For instance, often outlier data points are filtered prior to period estimation. However, such filtering can often remove potentially useful information. In contrast to this, \glspl{NN} do not directly remove sources of noise. Instead, they learn to identify such noise, while retaining signal within it, and incorporate this information into their predictions about the likelihood of a \gls{LC} containing the desired signal, in this case, a short-period signal. This approach makes them effective at processing noisy data~\citep{dong2014learning,hinton2012deep,xu2014deep}. For instance, if outliers occur due to a temporary systematic offset within the \gls{LC}, as is common in \gls{TESS} data, but still carry the periodic signal, the relevant information for detecting the periodic signal can be retained by the \gls{NN} where a simple omission of outliers would remove it.

As a generalized function approximator, a sufficiently large \gls{NN} has the potential to learn any hand-crafted function arbitrarily well, as is shown by the Universal Approximation Theorem~\citep{cybenko1989approximation,leshno1993multilayer,zhou2020universality}. This includes functions such at those used to produce a periodogram for a \gls{LC}. Furthermore, if improvements can be made to the hand-crafted transformation that yields improved results, a \gls{NN} has the potential to learn that improved function instead. This inherent property of \glspl{NN} gives them the potential to outperform their hand-crafted counterparts.

The computational speed of the method is another advantage of using \glspl{NN} in detecting short-period signals, as we will show from the computational performance of our method.

\subsection{Short-period main sequence binaries background}\label{subsec:binaries_background}

The first of the two primary populations identified by our \gls{NN} are short-period \gls{MS} binaries, which, at the shortest periods, consist of red dwarf stars. Due to their low mass, red dwarf stars burn their nuclear fuel at a slow rate, and every red dwarf star that has ever formed is still on the \gls{MS}. Although red dwarf stars are the most common type of star in the galaxy~\citep{bochanski2010luminosity}, how they evolve in close binary systems, including in contact binaries, is not well understood. There is a dramatic cut-off of such binaries at \approximately0.22 days~\cite{rucinski1992can,rucinski2007short}. Notably, the reason for this cut-off is still under investigation. The cut-off was originally often attributed to the stars reaching their fully convective limit~\citep{rucinski1992can}. More recently, magnetic braking~\citep{stepien2007low} and unstable mass transfer~\citep{jiang2012short} have been implicated of playing a prominent role. At any rate, a larger sample size may help to more clearly define the evolutionary and formation processes of these and other short-period \gls{MS} binaries.
 \subsection{Delta Scuti background}\label{subsec:delta_scuti_background}

\gls{DS} stars are a type of pulsating variable star with intermediate mass, 1.5–2.5 solar
masses~\citep{bedding2020very}, located at the intersection of the classical instability strip~\citep{dupret2004theoretical} and the \gls{MS} stars on the \gls{HR} diagram~\citep{handler2009delta,breger2000delta}. 
Typically of spectral types A to F, the \gls{DS} star class is composed of both Pop I and Pop II stars~\citep{1944ApJ...100..137B}.
These stars, characterized by a period range spanning from 0.02 days to 0.25 days, demonstrate multiperiodic luminosity variability and can present both radial and non-radial pulsations~\citep{breger2000delta}. 

Stellar pulsations serve as a valuable tool in investigating the internal structures of stars. The excitation mechanism underlying these pulsations involves the cyclic transfer of kinetic energy from the internal energy of the mixture of gas and radiation in the ionization zones, particularly rich in elements like helium and hydrogen. By identifying the natural frequencies of their pulsation modes, it is possible to compare these observations with theoretical models improving our understanding of stellar properties and their evolutionary processes~\citep{breger2000delta,bedding2020very}. Notably, \gls{DS} stars were among the first pulsators whose asteroseismic potential was recognized~\citep{handler2009delta}.

As stars evolve, they may cross the instability strip and may pulsate. They could be stars moving from the \gls{MS} to the giant branch or pre-\gls{MS} stars evolving to \gls{ZAMS}~\citep{breger2000delta, breger1972pre}. \gls{DS} stars have also been detected in diverse scenarios, ranging from eclipsing binaries~\citep{kahraman2023discovery} to stars populating other galaxies~\citep{mateo1998dwarf}. This broad range of detection highlights that \gls{DS} stars are not uncommon.

Beyond their use to provide insight into the internal structures of stars, \gls{DS} stars are often used as standard candles, due to their adherence to a period-luminosity relation in certain passbands.
      \section{Data}\label{sec:data}

\begin{figure*}
    \centering
    \begin{pycode}
        from data_resources.generate_representative_light_curve_figure import generate_representative_light_curve_and_folded_light_curve_figure0
        create_latex_figure_from_bokeh_layout(bokeh_layout=generate_representative_light_curve_and_folded_light_curve_figure0(),
                                              latex_figure_path='data_resources/representative_light_curve_and_folded_light_curve_figure0.png',
                                              latex_width=r'\textwidth',
                                              latex_height=r'0.8\textwidth')
    \end{pycode}
    \includegraphics[width=\textwidth]{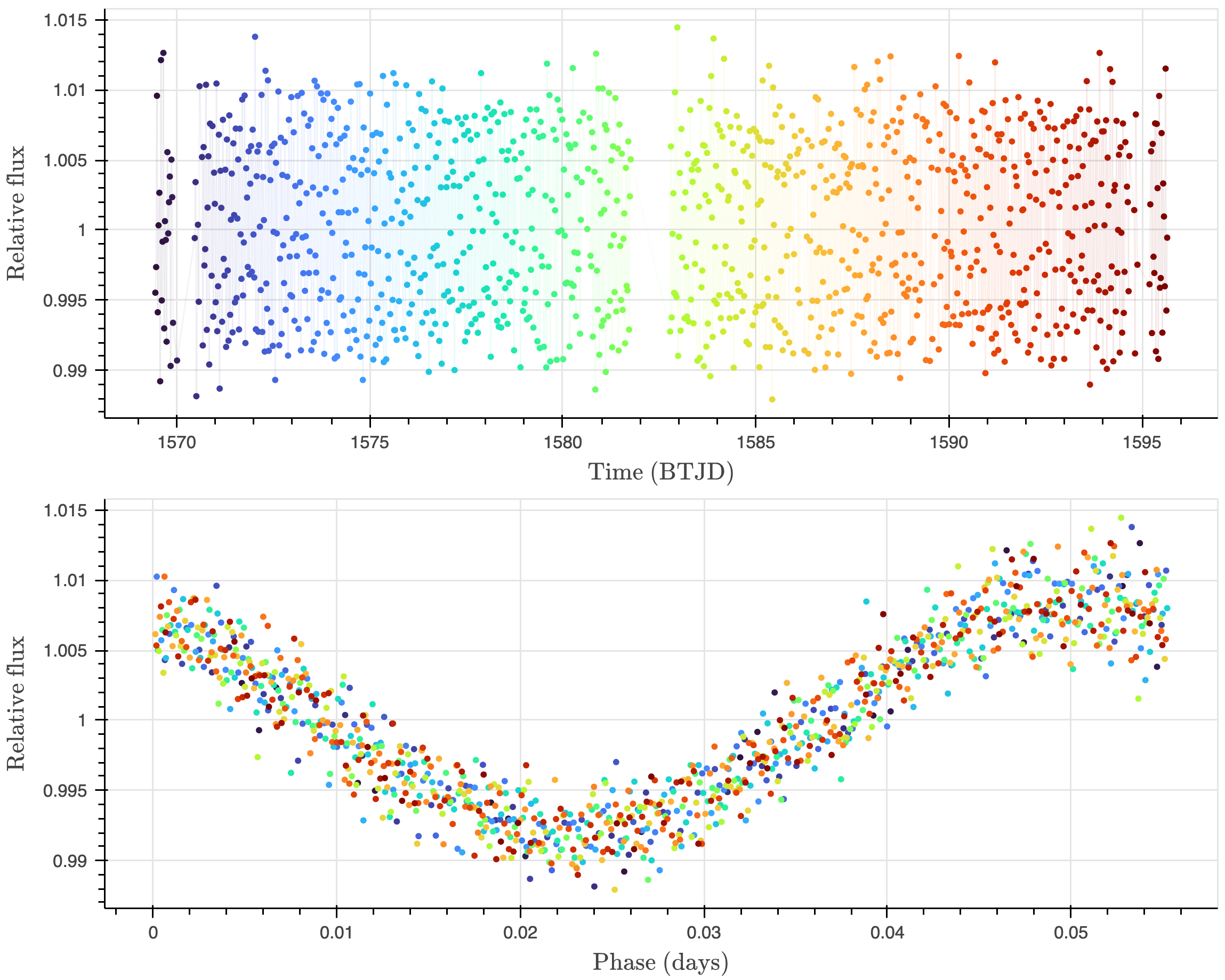}
    \caption{The FFI light curve and folded light curve for TIC ID \exampleLightCurveTicIdZero{} sector \exampleLightCurveSectorZero{}. This light curve was chosen as a typical example of the light curves identified by our NN. A notable aspect is that only 2 or 3 data points exist for each period (\exampleLightCurvePeriodInHoursZero{} hours), resulting in the periodicity not being clear to a human observer in the unfolded light curve. Despite no specific periodicity-detecting mechanisms being included in the NN, it learns to identify such periods in the unfolded data. The time is given in \gls{TESS} Barycentric Julian Day (BTJD) time~\citep{tenenbaum2018tess}. With the Julian Day in the Barycentric Dynamical Time standard (BJD), $\text{BTJD} = \text{BJD}-2457000.0$. BJD is used for its accurate time standard which accounts for many different timing corrections, including leap seconds~\citep[\eg,][]{eastman2010achieving}. The flux given is median normalized flux for the \gls{LC}. Color is based on unfolded time. This is an example from our \gls{DS} clusters (see \cref{subsec:partitioning_the_data} for clusters explanation).}
    \label{fig:representative_light_curve_figure0}
\end{figure*}

In this work, we use \gls{LC} data, \ie, measures of flux over time. \cref{fig:representative_light_curve_figure0} shows an example of a \gls{TESS} \gls{LC}, showing flux vs.~time of the \gls{TESS} target for TIC ID \exampleLightCurveTicIdZero{} sector \exampleLightCurveSectorZero{}. The \glspl{LC} were produced by the \texttt{eleanor} pipeline~\citep{feinstein2019eleanor} from raw flux measurements provided by \gls{TESS}.

\begin{figure*}
    \centering
    \begin{pycode}
        from data_resources.generate_representative_light_curve_figure import generate_representative_light_curve_and_folded_light_curve_figure1
        create_latex_figure_from_bokeh_layout(bokeh_layout=generate_representative_light_curve_and_folded_light_curve_figure1(),
                                              latex_figure_path='data_resources/representative_light_curve_and_folded_light_curve_figure1.png',
                                              latex_width=r'\textwidth',
                                              latex_height=r'0.8\textwidth')
    \end{pycode}
    \includegraphics[width=\textwidth]{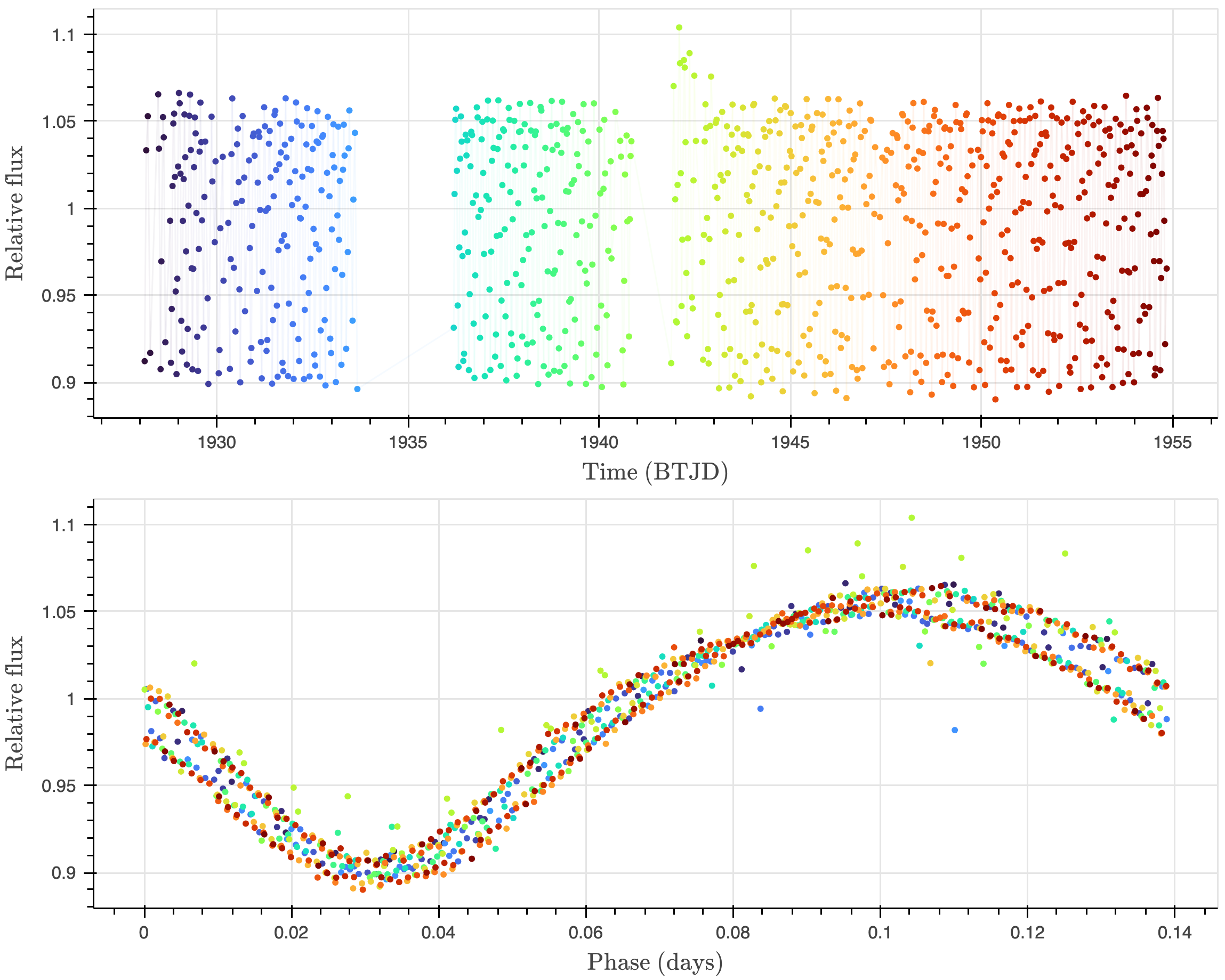}
    \caption{The FFI light curve and folded light curve for TIC ID \exampleLightCurveTicIdOne{} sector \exampleLightCurveSectorOne{}. This is an example from our \gls{MS} binary cluster with clearly distinct binary signal peaks (see \cref{subsec:partitioning_the_data} for clusters explanation). Presented the same as \cref{fig:representative_light_curve_figure0}. See \cref{fig:representative_light_curve_figure0} for details.}
    \label{fig:representative_light_curve_figure1}
\end{figure*}

The \gls{LC} shown in \cref{fig:representative_light_curve_figure0} is a typical example of a short-period variable identified by our \gls{NN}. There is a gap in the middle of the data caused by the spacecraft pausing its observing to downlink data to Earth~\citep{tenenbaum2018tess}. Notably, since we only searched for variables ranging from 1 hour to 5 hours, the \gls{FFI} cadence of 30 minutes results in very few data points per period. The \gls{LC} shown in \cref{fig:representative_light_curve_figure0} has a period of \exampleLightCurvePeriodInHoursZero{} hours, resulting in only 2 or 3 data points per period. This makes it difficult for a human to identify the periodicity in the unfolded \glspl{LC}. \cref{fig:representative_light_curve_figure0} shows an example from our \gls{DS} partitions (see \cref{subsec:partitioning_the_data} for partitioning explanation). \cref{fig:representative_light_curve_figure1} shows another example \gls{LC}, this time taken from our \gls{MS} binary partition.

Ideally, a \gls{LC} would contain only the flux from a single \gls{TESS} target (typically a star system). However, in reality each \gls{TESS} pixel covers \approximately21 arcseconds of the sky, and \gls{TESS}'s point spread function results in blending between pixel measurements. For these reasons, a LC will contain flux from multiple targets. This often makes it challenging to determine which source the signal (or noise) is originating from.

\Gls{TESS} takes measurements of a large portion of the sky at regular intervals. During the primary mission, this interval was every 2 minutes. However, due to limitations of the spacecraft's storage and downlinking capabilities, only a small portion of this 2-minute cadence data is stored and downlinked~\citep{ricker2014transiting}. However, at a 30-minute cadence, all pixels' accumulated values are downlinked to Earth. These \glspl{FFI} cover a much larger number of targets at a lower time resolution~\citep{tenenbaum2018tess}. We used \approximately67 million 30-minute cadence \glspl{LC} (with \gls{TESS} magnitudes \textless15) in this work, and this is the primary data set investigated by our \gls{NN}.

\subsection{\Glsfmtlong{FFI} \glsfmtlong{LC} production}
For details of the \gls{FFI} \gls{LC} production, see \citet{powell2022ffi}. Briefly, \citet{powell2022ffi} used the 129,000-core \textit{Discover} supercomputer at the NASA Center for Climate Simulation, to build \gls{FFI} \glspl{LC} for all sources observed by \gls{TESS} down to 15th magnitude. All original and calibrated FFIs were produced by the \gls{TESS} Science Processing Operations Center~\citep{jenkins2016tess}. Target lists were created through a parallelized implementation of \texttt{tess-point}~\citep{burke2020tess} on the \gls{TIC} provided by the \citet{masttess}. The \glspl{LC} for each sector were constructed in 1--4 days of wall clock time (for a total of over 100 CPU-years), depending on the density of targets in the sector, through a parallelized implementation of the \texttt{eleanor} Python module~\citep{feinstein2019eleanor}. \approximately67 million \glspl{LC} were used for this work. As of this writing, the \glspl{LC} of the first 10 sectors of the \gls{TESS} primary mission data have been made publicly available by \citet{powell2022ffi}. We have used the full 26 sectors of the primary mission in this work. The public data release of the remaining primary mission sectors of the data by \citet{powell2022ffi} is still ongoing. These single sector \glspl{LC} are the input to the pre-processing and, subsequently, our \gls{NN}. From this dataset, we use the flux with the processing labeled as ``corrected'' by \citet{powell2022ffi}.
     \section{Neural network pipeline}\label{sec:neural-network-pipeline}
The \gls{NN} architecture and data preprocessing used in this work is the same as in \citet{olmschenk2021transit}. In that work, we developed a \gls{NN} to identify transiting planets in \gls{TESS} \gls{FFI} data. The \gls{NN} architecture and data preprocessing from \citet{olmschenk2021transit} were designed with the intent to be generalizable and many of the explanations of design choices given there directly translate to the use case of short-period variables as well. Where we feel the connection is not immediately intuitive, we include additional explanation below. Also provided in \citet{olmschenk2021transit} is a short primer to \glspl{NN} which we believe will be helpful for readers of this work who are not familiar with \glspl{NN}.

\subsection{Training data}\label{subsec:training_data}
When designing and training the network, we use 80\% (\approximately54M \glspl{LC}) of the available TIC IDs as the training data targets and use another 10\% (\approximately7M \glspl{LC}) as validation data targets. These validation data are data that are set aside that the network is not trained on. Instead, these data are used to evaluate the predictive performance of the network during the training process. This entails measuring the correctness of the network predictions on the validation data, which the network has not been trained with but where we know the correct answer. The remaining 10\% is reserved to be used as a test dataset for a future evaluation. The test data is intended to evaluate the trained network after all design decisions are finalized. Several of the specific network and training configuration decisions were guided by preliminary performance results on the validation data. However, this validation data evaluation and the test data evaluation are beyond the scope of this work. The synthetic data used for training was not designed to accurately reflect the true distribution of expected periodic signals. Instead, it was designed to be a dataset which would produce a trained \gls{NN} that can identify short-period targets. Additionally, we believe such as test evaluation without further context would be misleading. For example, if we made the synthetic dataset less challenging (e.g., by increasing the lower bounds of the amplitudes) the network would preform better on the test dataset, but perform worse on real data. Instead, we provide evaluations on real data below.

Our training data consists of four collections of \glspl{LC}. First, are the regular \gls{TESS} \glspl{LC}. These are treated as negative samples during training. Although there will be rare false negative samples in this collection, due to their rarity and the \gls{NN}'s statistical nature, they have minimal impact on the training process. The second collection is the same real \gls{TESS} \glspl{LC}, but injected with synthetic short periodic signals. The synthetic periodic signals are mixed sine and sawtooth signals. Due to the relatively small number of data points per period and the many variations of signals that can be produced with such a mixed signal, we found these synthetic signals to be sufficient to train our \gls{NN}. These are the positive examples used during training. The third collection is the same real \gls{TESS} \glspl{LC} injected with synthetic longer-period signals. We use this collection as a hard negative training case (a negative sample we expect to be challenging for the network) as it prevents the \gls{NN} from fitting some artifact of the data (e.g., looking for a perfect sawtooth wave existing in the signal) and instead requires the network focus on the feature we are interested in (e.g., short periodicity). The fourth collection is the regular \glspl{LC} injected with uniform noise with an amplitude profile that matches those in the synthetic periodic signals. This is to prevent the network from trying to learn that an increased variance in values from an injected signal should be used as an indicator of a positive sample. Samples from the four collections are used in equal ratios for each batch of data during training. During the training process, for each training sample with a synthetic signal injected, a new random signal was generated on-the-fly, resulting in effectively infinite training signals.

\begin{figure}
    \centering
    \begin{pycode}
        from pipeline_resources.generate_synthetic_light_curves_figure import generate_synthetic_light_curves_figure
        create_latex_figure_from_bokeh_layout(bokeh_layout=generate_synthetic_light_curves_figure(),
                                              latex_figure_path='pipeline_resources/synthetic_light_curves_figure.png',
                                              latex_width=r'\columnwidth',
                                              latex_height=r'0.8\columnwidth')
    \end{pycode}
    \includegraphics[width=\columnwidth]{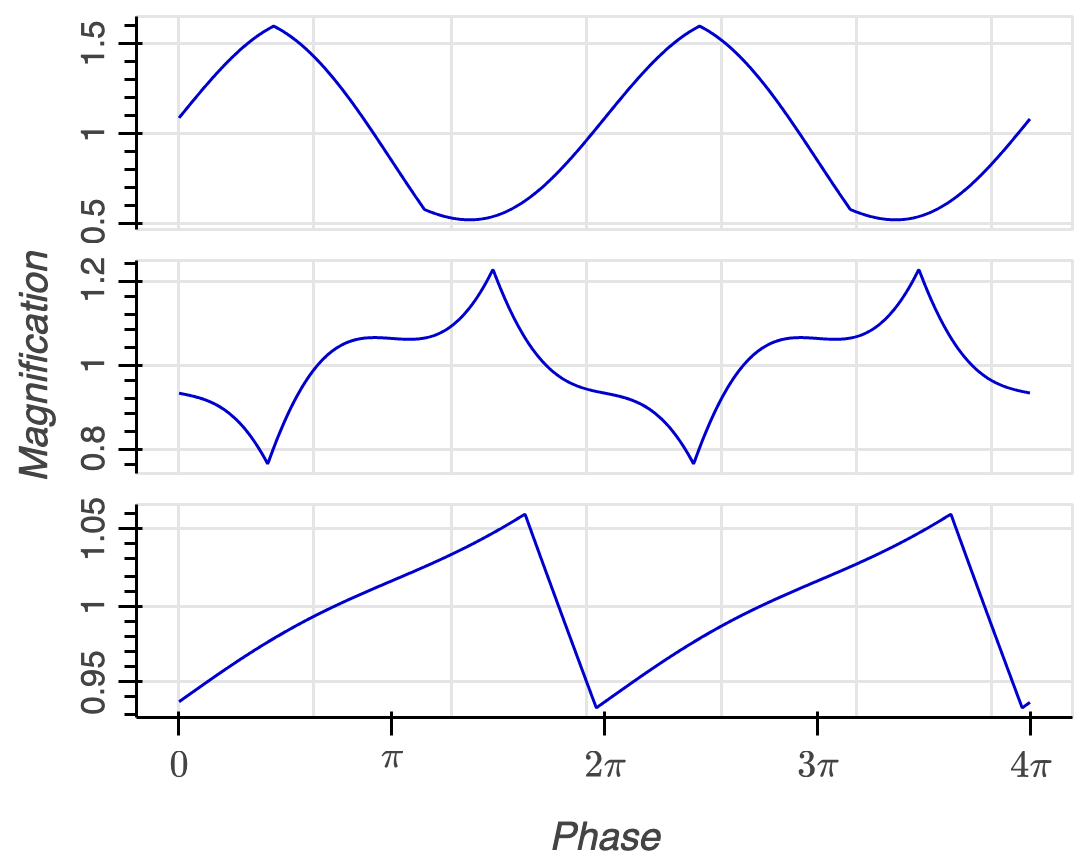}
    \caption{A random set of examples of synthetic periodic signals to be injected into TESS light curves as training data. Two periods of each synthetic signal are shown.}
    \label{fig:synthetic_light_curves_figure}
\end{figure}

The following are the specifics of the synthetic periodic signals we generated. A period is selected from $\mathcal{U}(0.25, 5)$ hours for the short-period signals and from $\mathcal{U}(9, 20)$ hours for the longer-period signals. Each signal had a relative amplitude selected from $\mathcal{U}(0.001, 1)$ independently for both the sine and sawtooth component. A random phase from $\mathcal{U}(0, 2\pi)$ was selected to offset the sine and sawtooth components. The fraction of the sawtooth cycle which consisted of the rising ramp was selected from $\mathcal{U}(0, 1)$ (with the remainder being the falling ramp). The signal was injected at a random phase into the real \gls{TESS} \gls{LC}. A random sample of these synthetic \glspl{LC} are shown in \cref{fig:synthetic_light_curves_figure}.

\subsection{Network architecture}\label{subsec:network_architecture}
\begin{figure*}
    \centering
    \begin{minipage}{0.4\textwidth}
        \centering
        \subfloat[The full network.\label{fig:ffi_hades_network}]{
            \centering
            \includegraphics[width=0.85\textwidth]{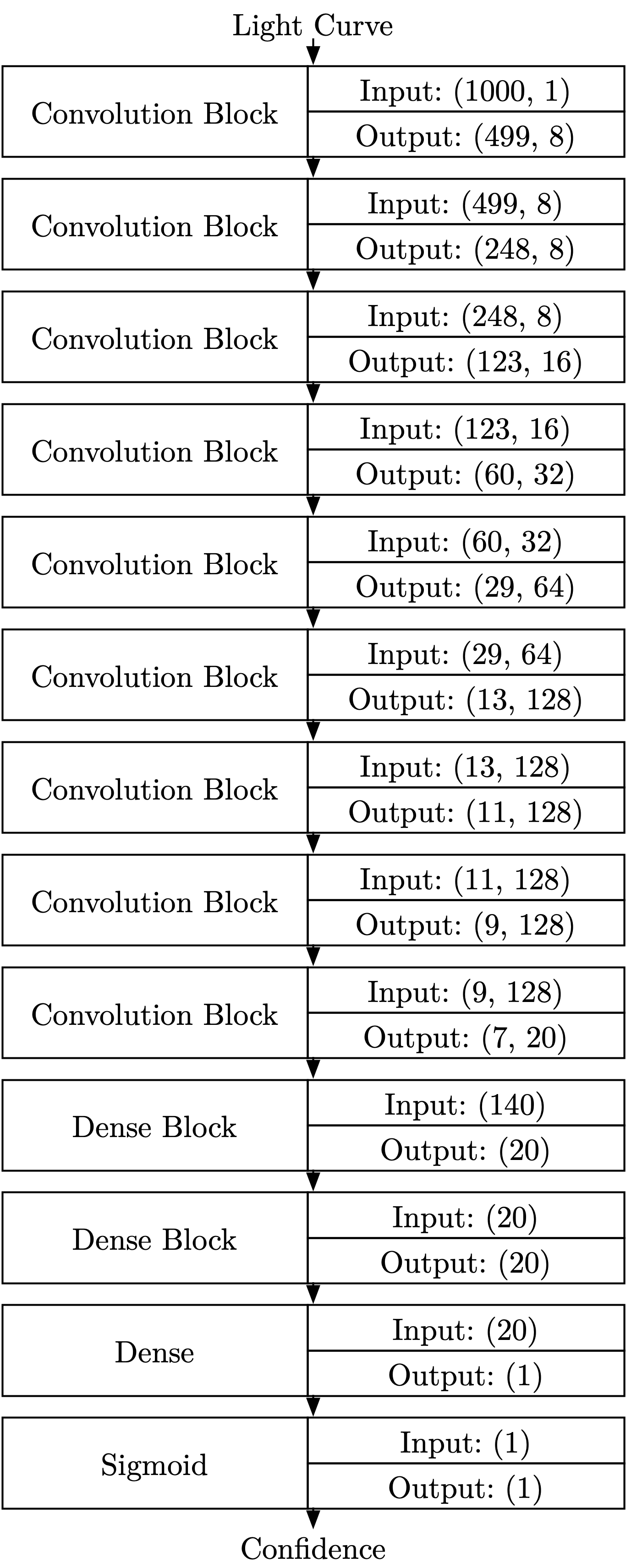}
        }
    \end{minipage}
    \begin{minipage}{0.25\textwidth}
        \vfill
        \centering
        \subfloat[The outline of a convolution block structure.\label{fig:ffi_hades_convolution_block}]{
            \centering
            \includegraphics[width=0.65\textwidth]{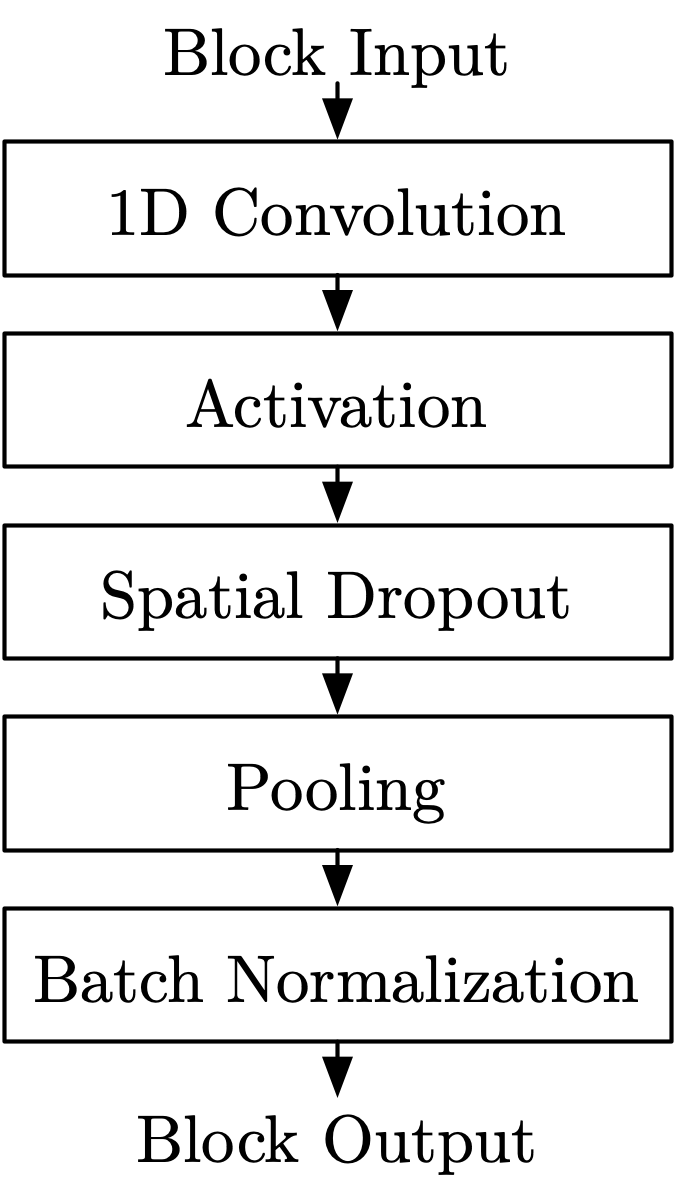}
        }
        \\
        \vspace{1cm}
        \subfloat[The outline of a dense block structure.\label{fig:ffi_hades_dense_block}]{
            \centering
            \includegraphics[width=0.65\textwidth]{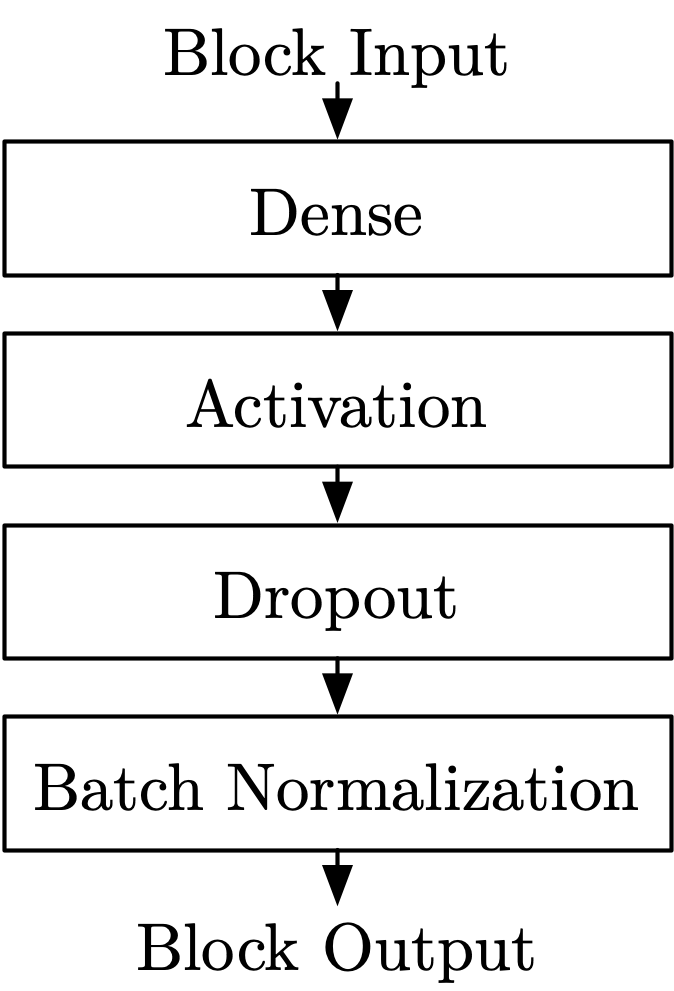}
        }
        \vfill
    \end{minipage}
    \caption{An overview of the architecture of the convolutional neural network used in this work. See \citet{olmschenk2021transit}
    for details. All convolution/dense layers within a block use a number of filters/units equivalent to the size of
    the last dimension of their output tensor. A kernel size of 3 is used in all convolutional layers. For clarity of
    the diagram, three deviations from the diagram blocks are not shown. First, the first convolution block and the last
    dense block do not apply dropout or batch normalization. Second, the final convolution block uses a standard
    dropout instead of spatial dropout as the following layer is a dense layer. Third, pooling is only used by the first 6
    convolution blocks. The remaining convolution blocks do not use pooling.}
    \label{fig:network_diagram}
\end{figure*}

In this work, we use a 1D \gls{CNN} that was originally designed and developed in \citet{olmschenk2021transit}. This \gls{NN} architecture is shown in \cref{fig:network_diagram}.
 Refer to \citet{olmschenk2021transit} for the details of the \gls{CNN} design.

Our \gls{NN} framework code is available online\footnote{\url{https://github.com/golmschenk/ramjet} (see \url{https://github.com/golmschenk/ramjet/releases/tag/short_period_variable_neural_network_paper} for the code version used in this work)}. This framework is also installable as a PyPI package\footnote{\url{https://pypi.org/project/astroramjet/}}. Documentation for the \gls{NN} framework is also available online \footnote{\url{https://astroramjet.readthedocs.io/en/latest/}}. This \gls{NN} framework is a generalized photometric \gls{NN} framework. Our code that applies this generalized framework to the specific task of short-period variable identification can be found online\footnote{\url{https://github.com/golmschenk/generalized-photometric-neural-network-experiments}}.

Similar to the situation in \citet{olmschenk2021transit}, the 1D \gls{CNN} is an apt choice for searching for short-period variables due to how it constructs high-level global features from low-level local features. First, we expect the \gls{NN} to find features such as individual peaks and troughs as low-level features. In early layers, the \gls{NN} will likely ignore the positions of these features in the \gls{LC}, and only determine their presence based on the local \gls{LC} shape. As such, the early layers of our network are convolutional layers, which treat each segment of the \gls{LC} identically~\citep{krizhevsky2012imagenet}, \eg, they search each portion of the LC for a peak or trough occurring in that location. Only after local level features, \eg, individual wave cycles, are discovered do we expect the network to combine these features into global level features---in this case repeating periodic wave cycles. The other advantages of a \gls{CNN}, such as its prevention of overfitting, are the same as explained in \citet{olmschenk2021transit}.

No prior short-period variable parameter information is required by our \gls{NN} (\eg, no phase folding or other prior information extraction is performed). The only inputs to the \gls{NN} are the fluxes of the \glspl{LC}. Inference on a \gls{LC} is performed by the \gls{NN} in \approximately5ms on a single GPU. This allows for inference of the entire \approximately67M \gls{FFI} \gls{LC} dataset to be completed in a few days. The network training took \approximately5 days on a single GPU, with this training time able to be decreased approximately linearly with number of GPUs used in parallel.

The specific number of layers and size of each layer (as seen in \cref{fig:network_diagram}) was decided through limited experimentation, vaguely guided by prior experience with network structures in computer vision. While a systematic search of network structures is beyond the scope of this work, we note that a minor change to our \gls{NN} (\eg, adding/removing a layer), while adjusting the remainder of the \gls{NN} to produce the same output size, does not produce a trained \gls{NN} that produces extremely different results.

\subsection{Pre-processing}\label{subsec:pre-processing}
We use the same pre-processing of the \glspl{LC} as in \citet{olmschenk2021transit}. The primary purpose of this pre-processing is to prevent the network from overfitting and to encourage generalization of learned features. Briefly, during the training phase, a random data point removal is applied and a random rolling of the data is applied. During both the training phase and the inference phase, a uniform lengthening is applied and a modified z-score normalization is applied. See \citet{olmschenk2021transit} for details on these pre-processing methods and why they are applied.

\subsection{Post-neural network processing}\label{sec:post_neural_network_processing}
After the \gls{NN} has identified likely short-period variable candidates, we run a process on the candidates to determine the periodicity and remove false positives. Where the \gls{NN} only used the flux values of the \gls{LC}, here we use additional external information to remove false positives from the \gls{NN} filtered candidate list. The complete code used to perform this processing can be found at \url{https://github.com/golmschenk/generalized-photometric-neural-network-experiments}. Below we describe the most salient points.

First, we selected the top 50,000 \glspl{LC} given the highest confidence score from the \gls{NN}. The number 50,000 is arbitrary. A larger sample with lower confidences could be selected, but we opted for a relatively high confidence cutoff.

Then, we estimated candidate periodicity using a Lomb-Scargle periodogram. The Lomb-Scargle periodogram search was limited to a period of 1 hour (twice the sampling rate of \gls{TESS} \glspl{FFI}) up to 10 days. From the resulting periodogram, of the frequencies with a power within 10\% of the max power, we take the highest frequency, then find the local power maximum from that frequency. We take this frequency/period as the short-period variable's frequency/period. Any targets with periods greater than 5 hours are discarded. These steps are chosen to reduce the chances of choosing a shorter period alias. Here, we also fold the light curve, and, binning into 25 bins in phase space, determine the minimum and maximum bins of the light curve based on the median value of the bin.

Next, we remove false positives which have photometric centers of variability that do not align with the target. This is done to prevent cases where a nearby variable is the real source of the variability. To accomplish this, we first use TESScut~\citep{brasseur2019astrocut} to obtain the time-series raw image data of the pixels surrounding the target. We use an image of $10\times10$ centered on the target. The time-series image data is folded on the period determined previously, and again placed into the 25 bins across phase space determined previously. Those which fall into the minimum and maximum bins of the previous step are used for the following variability photometric centroid estimation. We find median values of the binned images for the minimum and maximum bins, then take the difference of these resulting values. The centroid of these values is compared to the estimated position of the target from the \gls{TIC}. If the estimated centroid is separated from the target position by more than 21 arcseconds (the angular size of a side of a TESS pixel), the candidate is discarded.

The amplitude is estimated as half of the difference between the maximum and minimum of the above binning. The relative amplitude is this previous amplitude divided by the median flux. The version of the relative amplitude with contamination uses the median flux of the original light curve. The version of the relative amplitude without contamination scales this value to take into account the estimated contamination ratio provided via the \gls{TIC} where available.

Additional target stellar properties are taken from the \gls{TIC} and \gls{Gaia} where available.
     \section{Results}\label{sec:results}

The table of our identified short-period variable candidates is available as a supplemental file to this work. \cref{tab:short_period_variables_table} shows a random\footnote{To prevent the presentation of unrepresentative data, any references to a ``random" selection of results to be presented in this work involve sorting the data in some logical way, performing a seed 0 shuffle, then selecting the shuffled elements in order. In this case, sorted by TIC ID then shuffled with seed 0.} sample of 100 rows from the complete table.

\begin{figure}
    \centering
    \begin{pycode}
        from results_resources.period_distribution_figure import generate_period_distribution_figure
        create_latex_figure_from_bokeh_figure(bokeh_figure=generate_period_distribution_figure(),
                                              latex_figure_path='results_resources/period_distribution_figure.png',
                                              latex_width='\columnwidth',
                                              latex_height='\columnwidth')
    \end{pycode}
    \includegraphics[width=\columnwidth]{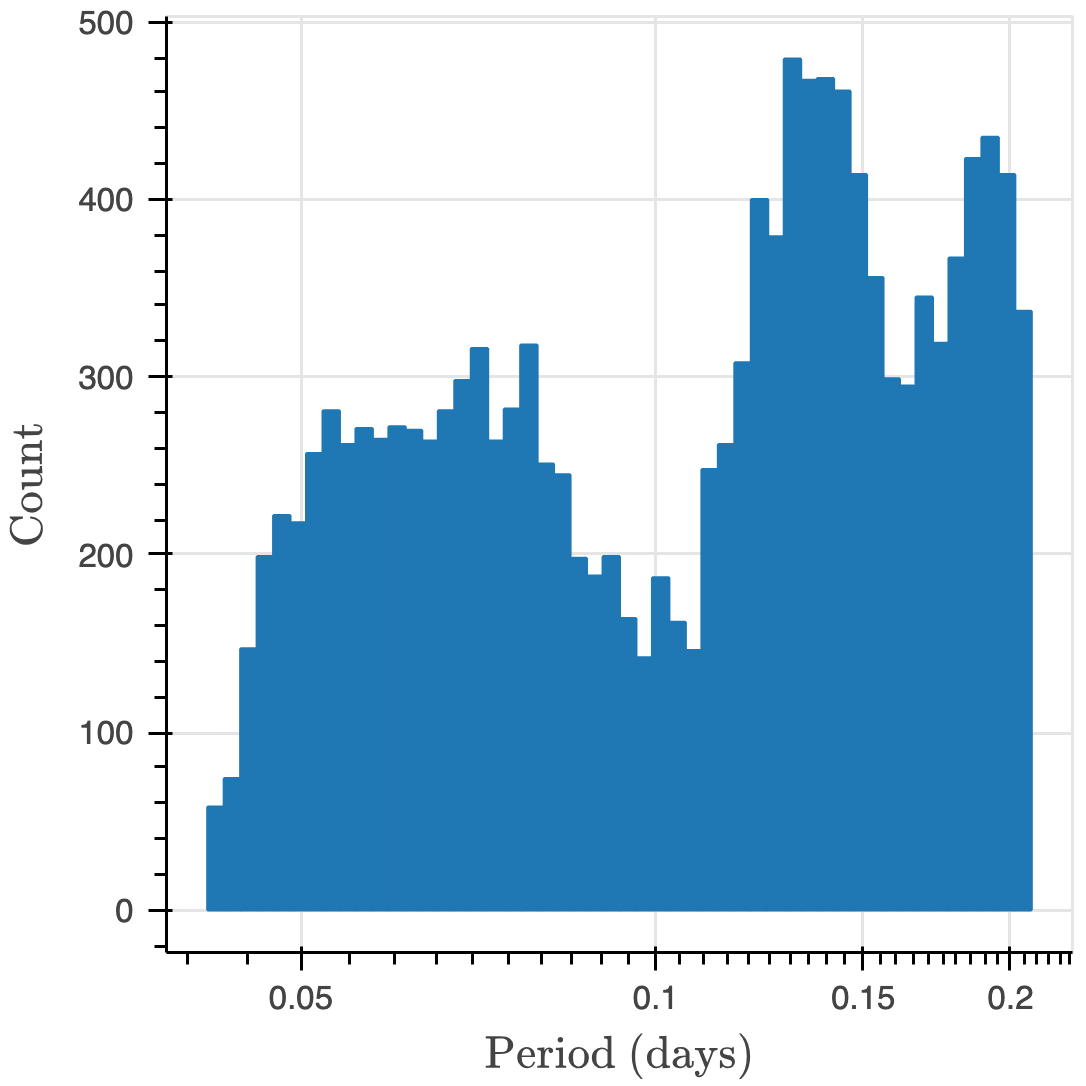}
    \caption{The period distribution of the short-period variables as estimated by our pipeline. Note that for binary systems, this
    estimated period may be half of the system's full orbital period.}
    \label{fig:period_distribution_figure}
\end{figure}

\begin{figure}
    \centering
    \begin{pycode}
        from results_resources.period_vs_temperature_density_figure import generate_period_vs_temperature_density_figure
        create_latex_figure_from_bokeh_figure(bokeh_figure=generate_period_vs_temperature_density_figure(),
                                              latex_figure_path='results_resources/period_vs_temperature_density_figure.png',
                                              latex_width='\columnwidth',
                                              latex_height='\columnwidth')
    \end{pycode}
    \includegraphics[width=\columnwidth]{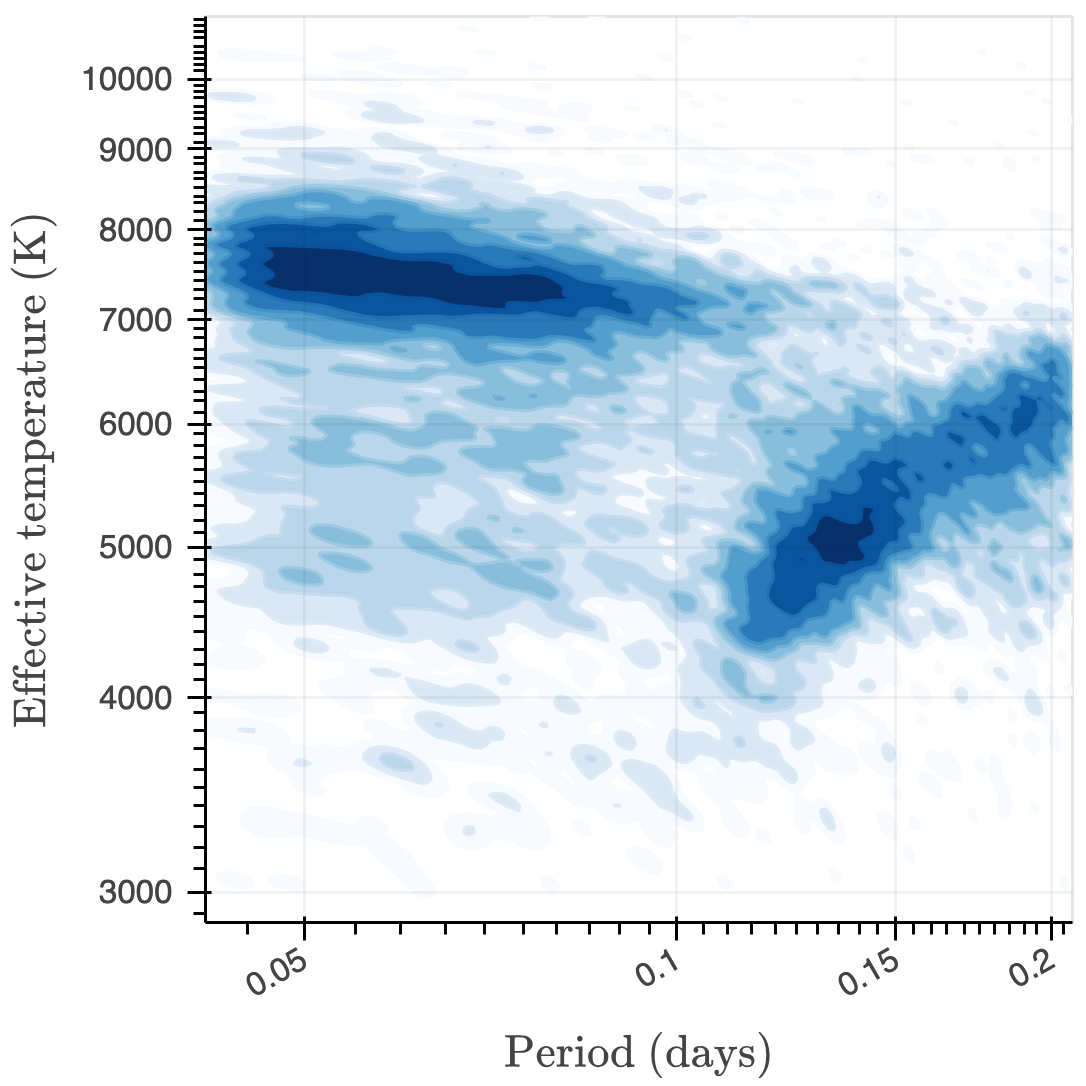}
    \caption{A kernel density estimation of effective temperature vs period of the short-period variables. The effective temperature of the targets is taken from the TIC.}
    \label{fig:period_vs_temperature_density_figure}
\end{figure}

\cref{fig:period_distribution_figure} shows the distribution of the periods of the candidates. These periods are compared against the effective temperatures of the short-period variables in \cref{fig:period_vs_temperature_density_figure}. The longer period population corresponds to \gls{MS} binaries while the shorter period population corresponds to \gls{DS}. These were identified by comparing the population properties against known populations~\citet{ziaali2019period,barac2022revisiting,Fetherolf_2023}. Throughout the rest of the results and analysis, we have assumed these two populations to primarily contain \gls{MS} binaries and \gls{DS} respectively. However, we emphasize that we have not performed any form of detailed modeling on these targets, and these two populations almost certainly contain many targets which are not directly from these two classes of objects. The \gls{DS} population itself can be seen to be separated into two populations when comparing the period to the absolute magnitude (\cref{fig:absolute_v_magnitude_vs_log_period_clusters_colored_figure}).

\subsection{Human vetting}\label{subsec:human_vetting}
To help confirm our \gls{NN} accomplished its goal of identifying short-period variable targets, we inspected 500 random \glspl{LC} output by our pipeline. We visualized these \glspl{LC} folded on the periods provided by our pipeline. Of the 500, 492 had the periodic signal immediately obvious in the folded version of the \glspl{LC}. Of the remaining 8, 6 had a low signal-to-noise ratio. In all 6 cases, a median binned version of the phase folded light curve presented apparent trough and peak portion of the signal. While it's reasonably likely these are true signals, our post-processing also uses such a binning process to remove false positives. So it is possible that, given enough random binned data, these apparent signals may simply be the only kind of random noise which will pass through the post-processing. The other 2 of the 8 mentioned above had a period very near 1 hour (within 0.25 seconds) which, given the 30 minute cadence of the data, resulted in all the data points appearing near only two phases in the folded \gls{LC} with no data points in between. In both of these cases, the two phase segments with data points were consistent with a periodic signal (with one phase segment sloping upward and the other sloping downward), however, we feel there is too little phase coverage to be confident of a periodic signal in these cases. With this, we are confident 492 of the 500 (98.4\%) of the randomly selected \glspl{LC} demonstrate a short-period signal. Of the remaining 1.6\%, all have some indication of a periodic signal, but they are less certain. \subsection{Evaluation against an existing catalog}\label{subsec:evaluation_against_existing_catalog}
We preform an evaluation using variable targets identified by \citet{Fetherolf_2023}. \citet{Fetherolf_2023} presents a catalog of variables stars identified in \gls{TESS} 2-minute cadence \glspl{LC}. Here, along with evaluating the entire pipeline, we evaluate the predictive performance of the \gls{NN} alone to give a sense of the performance of the \gls{NN} by itself. We note that the post-\gls{NN}-processing is designed to remove false positives that the \gls{NN} mislabels, but it is valuable to still have an understanding of the predictive performance of the \gls{NN} alone.

We note that this evaluation is imperfect, as the targets for which 2-minute cadence data exists are typically much brighter than our average target in the FFI dataset. However, it provides at least understandable metric to gauge our results by. \citet{Fetherolf_2023} identified 5449 targets with periods between 1 and 5 hours. With repeats of these targets across sectors, we have 13,647 \gls{FFI} \glspl{LC} for these targets in our dataset. For this evaluation, we consider this to be the positive dataset. To make the comparison interpretable, we selected an equal number of negative \glspl{LC} for evaluation. We selected these negative \glspl{LC} randomly from the remaining targets with 2-minute cadence data available in the \gls{TESS} primary mission data. To be clear, we do not use 2-minute cadence data in our pipeline in any fashion. We only selected \glspl{LC} for targets for which 2-minute cadence exists to provide a collection of \glspl{LC} which have similar properties to those in the positive dataset (\eg, magnitude, contamination). We then perform inference on the \gls{FFI} \glspl{LC} of these targets with our \gls{NN}.

As noted elsewhere, the goal of our pipeline is not to find all short-period variables, but instead to provide a large collection of high confidence candidates. As such, in our main results, we only took the highest confidence candidates from our \gls{NN} with an arbitrarily cut-off of 50,000 \glspl{LC}. This corresponds to a confidence threshold of 0.99796987 from the \gls{NN}. It's important to note, that this is an uncalibrated confidence and has no explicit bearing on the expected probability of a \gls{LC} containing a short-period signal. In our use case, only the confidences relative to one another are relevant and the absolute confidence values have little meaning beyond being a cutoff threshold. We use this same confidence threshold during this evaluation. Other thresholds of confidence could be chosen for other use cases (\eg, attempting a complete survey).

Again, since the goal is not survey completion, but rather the identification of a large number of candidates, the metrics of interest in this evaluation are the \gls{TPR} and the \gls{FPR}. With our high confidence threshold value, 1,363 of the 13,647 positive samples are above the threshold. Assuming all the positive samples from \citet{Fetherolf_2023} are correct, this results in a 9.987\% \gls{TPR}. For the negative case, 16 of the 13,647 are above the threshold. However, upon inspecting these 16 \glspl{LC}, we find 11 of them have clear short-period signals from 1 to 5 hours in period, and they are actually positive samples. This means that only 5 of the negatives are mislabeled by the \gls{NN} resulting in a 0.03663\% \gls{FPR}. All 5 of these negative samples are discarded by the post-\gls{NN}-processing of our pipeline meaning that, at least for this evaluation sample, our pipeline admitted no false positives.

While our pipeline produced a perfect result on this evaluation dataset (producing only true positives and no false positives), it is important to note that this dataset is not perfectly representative of the data the pipeline was applied to in the real inference case. First, in the real-world case, the dataset is expected to be highly imbalanced with the vast majority of targets not exhibiting a 1 to 5 hour short-period signal. Second, the targets our \gls{NN} were trained on typically had a dimmer magnitude on average than the collection of targets in this evaluation. As this distribution of samples is not the same as those the network was trained with, the network may be providing either higher or lower confidences for negative samples than it would on a distribution that matches the full inference dataset (which could result in either increased or decreased predictive performance). So, while this evaluation provides some sense of \gls{NN}'s performance, it is likely the above human vetting process is more representative of the quality of the short-period variables identified in this work. \subsection{Partitioning the data}\label{subsec:partitioning_the_data}
\begin{figure}
    \centering
    \begin{pycode}
        from results_resources.absolute_v_magnitude_vs_log_period_clusters_colored_figure import generate_absolute_v_magnitude_vs_log_period_clusters_colored_figure
        create_latex_figure_from_bokeh_figure(bokeh_figure=generate_absolute_v_magnitude_vs_log_period_clusters_colored_figure(),
                                              latex_figure_path='results_resources/absolute_v_magnitude_vs_log_period_clusters_colored_figure.png',
                                              latex_width='\columnwidth',
                                              latex_height='\columnwidth')
    \end{pycode}
    \includegraphics[width=\columnwidth]{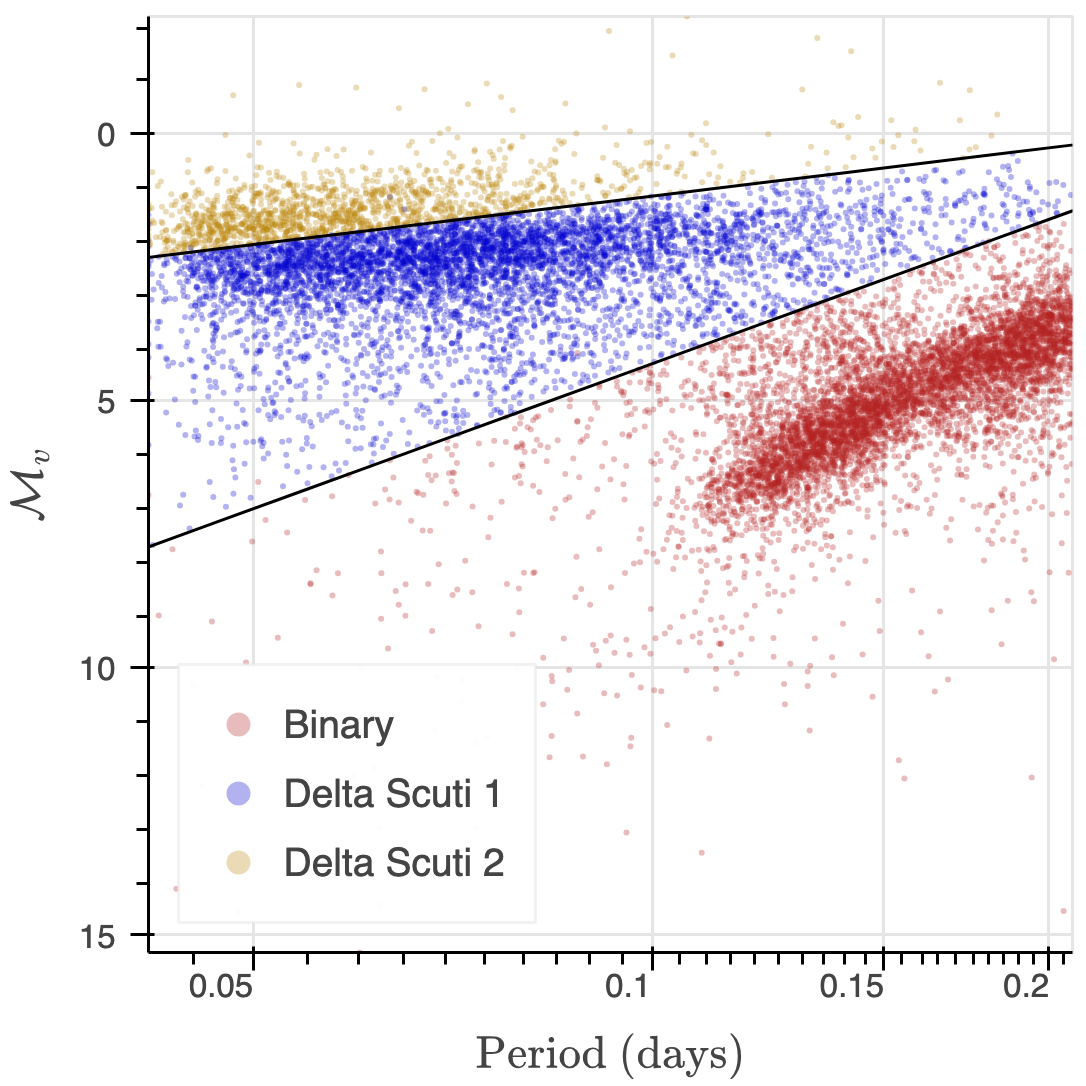}
    \caption{The absolute V magnitude v the period of the short-period variables. Note, these are the absolute V magnitudes provided by the TIC which does not
    assume the targets are binaries. Partition colors and partition lines are chosen based on a partitioning in log luminosity vs log period relations described in \cref{subsec:partitioning_the_data}.}
    \label{fig:absolute_v_magnitude_vs_log_period_clusters_colored_figure}
\end{figure}

In the absolute Johnson V magnitude vs log period distribution, we have defined linear partitions to divide the three primary populations of data, splitting the data into a \gls{MS} binary partition, a primary ridge \gls{DS} partition, and a second ridge \gls{DS} partition (\cref{fig:absolute_v_magnitude_vs_log_period_clusters_colored_figure}). We note, our \gls{NN} is not performing a classification. These populations of short-period variables were found during post-pipeline analysis of the properties of the targets that were identified by the \gls{NN}. We have partitioned the results based on literature values. Notably, the partition line between the primary ridge and secondary ridge \gls{DS} populations is set to the relation provided by \citet{barac2022revisiting}.

We additionally, attempted several simple fitting metrics to determine the partitions. However, due to outliers, unknown detection efficiencies, and indistinct boundaries (notably between primary and second ridge \gls{DS}), these simple metrics typically produced spurious partitions. Given that we have performed no detailed modeling of these different classes of objects and that these populations certainly contain other types of objects, we have opted to use the partitions based on literature values rather than produce more convoluted fitting metrics. However, we emphasize that these partitions should only be used for qualitative insights into the data and caution should be used when considering these partitions for any form of quantitative analysis. We also note that several smaller populations of interest likely exist in the data beyond these three, such as the small number of very high effective temperature variables discussed in \cref{subsec:hot_targets_analysis}.

\begin{figure}
    \centering
    \begin{pycode}
        from results_resources.period_vs_radius_figure import generate_period_vs_radius_figure
        create_latex_figure_from_bokeh_figure(bokeh_figure=generate_period_vs_radius_figure(),
                                              latex_figure_path='results_resources/period_vs_radius_figure.png',
                                              latex_width='\columnwidth',
                                              latex_height='1.3\columnwidth')
    \end{pycode}
    \includegraphics[width=\columnwidth]{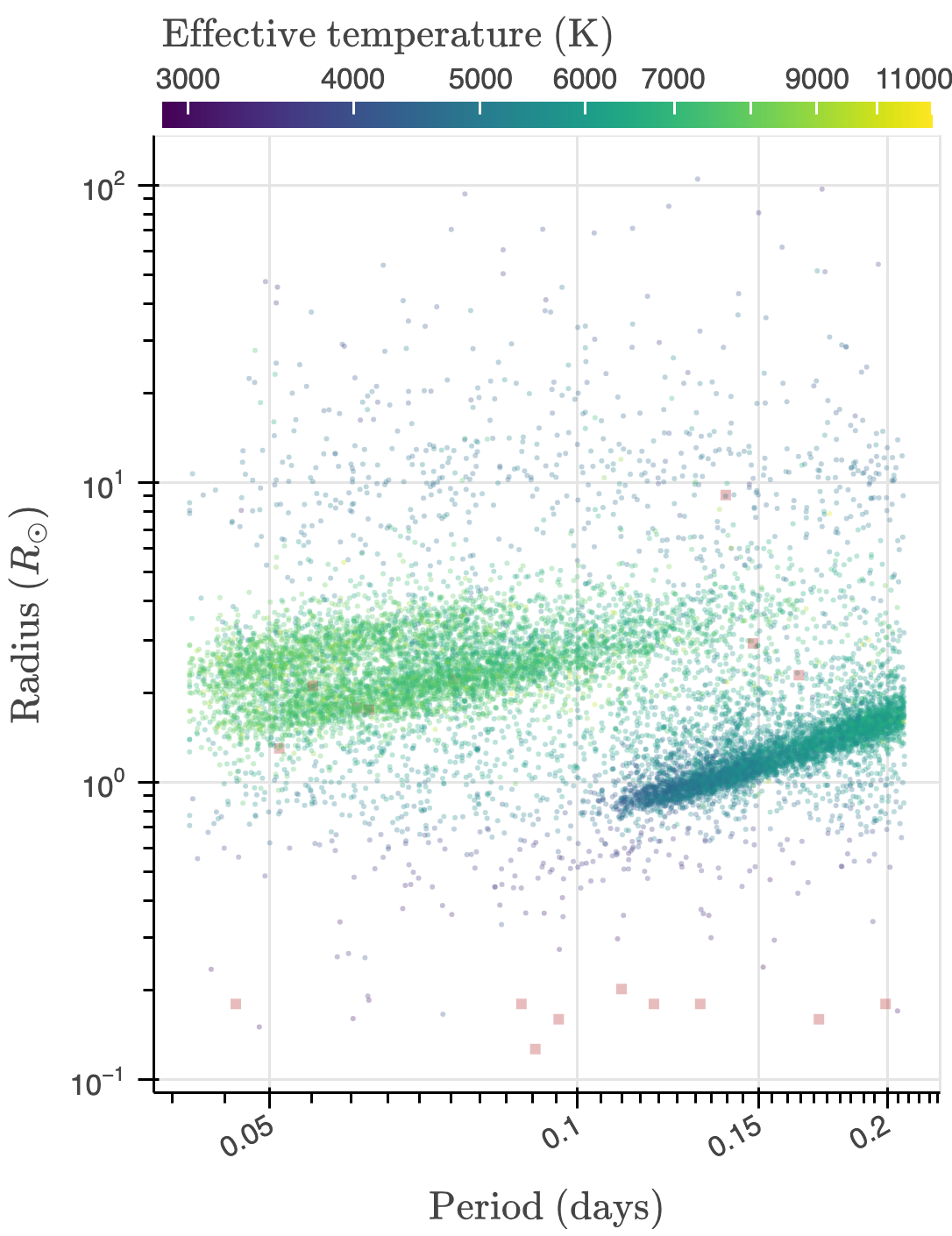}
    \caption{The radius vs the period of the short-period variables color by the variable effective temperature. Due to high effective temperature outliers, the upper limit of the effective temperature scale is restricted to \highTemperatureScaleLimit{}K and any targets above this effective temperature are shown as \highTemperatureColor{} squares. Note, caveats given in \cref{subsec:partitioning_the_data} about how these values are estimated and how the estimate value may differ from the true value, especially in regard to binaries.}
    \label{fig:period_vs_radius_figure}
\end{figure}

\cref{fig:absolute_v_magnitude_vs_log_period_clusters_colored_figure} shows absolute Johnson V magnitude vs period of the short-period variables along with the selected partition lines. The \gls{MS} binary partition targets are shown in \clusterColorRedDwarfBinary{}, the primary ridge \gls{DS} partition targets in \clusterColorDeltaScutiPrimaryRidge{}, and the second ridge \gls{DS} partition targets in \clusterColorDeltaScutiSecondRidge{}. These colors are consistent for any other figures using this clustering colorization, with the addition of \clusterColorNone{} for targets with unknown absolute Johnson V magnitude (which were then not included in the partitioning). Figures throughout this work will note when they use this coloring scheme. The primary and second ridges of the \gls{DS} are discussed in more detail in \cref{subsec:delta_scuti_analysis}. This separation is more clear in the radius vs period relation (\cref{fig:period_vs_radius_figure}), however, as the radius values are derived from the absolute magnitude values along with some assumptions of the target, we use the absolute magnitude vs the period relation to distinguish these clusters. Additionally, existing works typically use this absolute Johnson V magnitude vs period relation to define the ridges of the \gls{DS} targets.

Some caveats need to be considered when reviewing these results. First, the estimated period is the predominant period found during our processing explained above. Notably, for binaries, this estimated period will typically be half the orbital period. Second, except for the estimated periods, the properties of the targets listed here are taken from the \gls{TIC}. Some of the \gls{TIC} values may be taken from other existing catalogs or directly estimated through \gls{TESS} observations. However, when these values are not either directly observed or known from existing catalogs, as is often the case, they may be derived values. Notably, luminosity, absolute Johnson V magnitude, effective temperature, radii, and mass are derived from magnitude, color, and parallax. The details of how these values are derived can be found in \citet{stassun2018tess}. These derived values may cause some issues in the visualizations shown here. Most notably, these derived values are estimated for a single star, where in the case of binary systems this will result in a scale factor misrepresentation for various properties (e.g., radius).

As more detailed modeling of each of these targets with certainty is beyond the scope of this work, and we cannot know which of the targets do not actually belong in their assigned partitions, we present the original derived values. In any case where one of the properties of the target being plotted in a figure is unknown, that target will be excluded from the figure.

\begin{figure}
    \centering
    \begin{pycode}
        from results_resources.color_kde_clusters_colored_figure import generate_color_kde_clusters_colored_figure
        create_latex_figure_from_bokeh_figure(bokeh_figure=generate_color_kde_clusters_colored_figure(),
                                              latex_figure_path='results_resources/color_kde_clusters_colored_figure.png',
                                              latex_width='\columnwidth',
                                              latex_height='\columnwidth')
    \end{pycode}
    \includegraphics[width=\columnwidth]{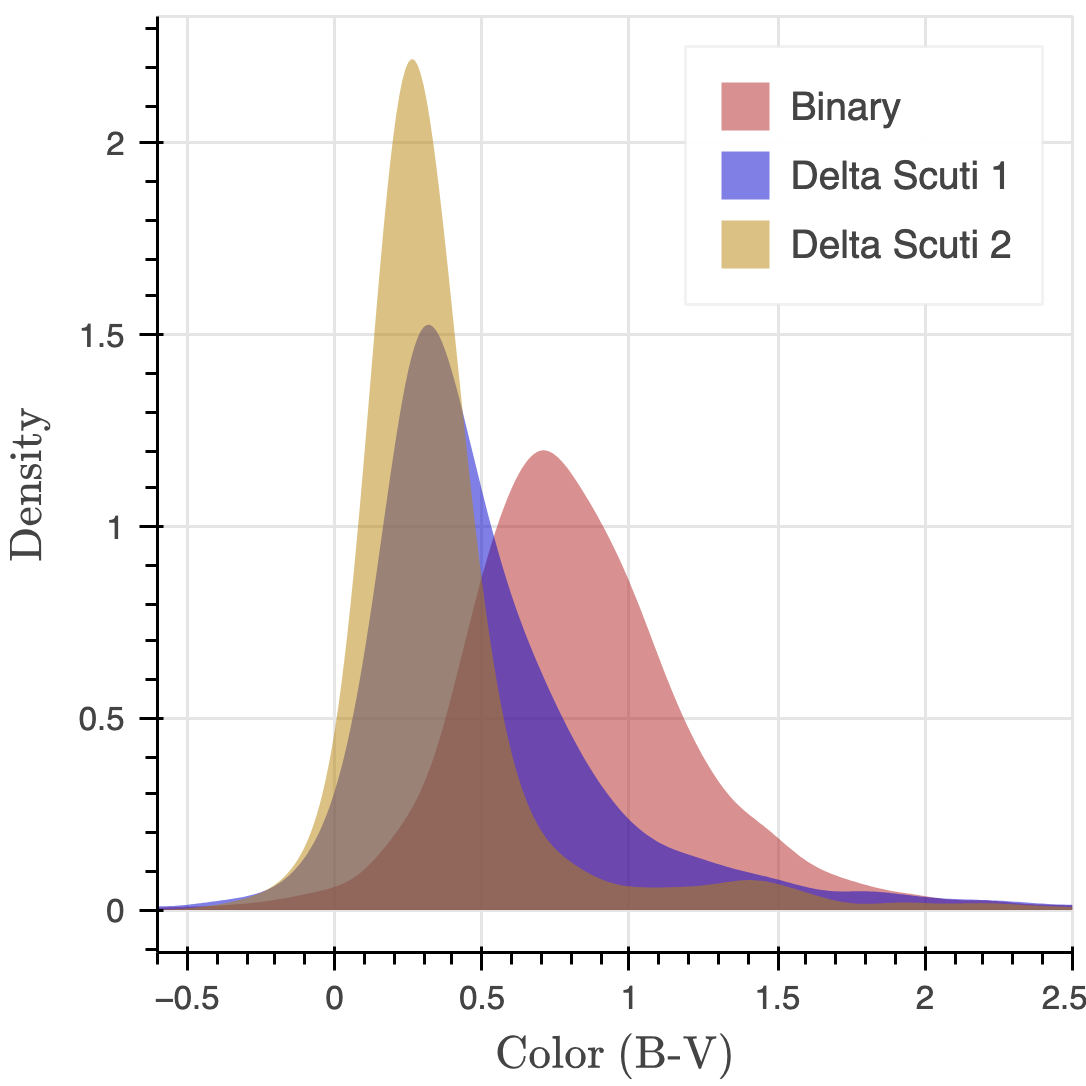}
    \caption{A kernel density estimation of the colors of the short-period variables. Color values come from the TIC. Distributions are colored by the partitions described in \cref{subsec:partitioning_the_data}.}
    \label{fig:color_kde_clusters_colored_figure}
\end{figure}

\begin{figure}
    \centering
    \begin{pycode}
        from results_resources.magnitude_kde_clusters_colored_figure import generate_magnitude_kde_clusters_colored_figure
        create_latex_figure_from_bokeh_figure(bokeh_figure=generate_magnitude_kde_clusters_colored_figure(),
                                              latex_figure_path='results_resources/magnitude_kde_clusters_colored_figure.png',
                                              latex_width='\columnwidth',
                                              latex_height='\columnwidth')
    \end{pycode}
    \includegraphics[width=\columnwidth]{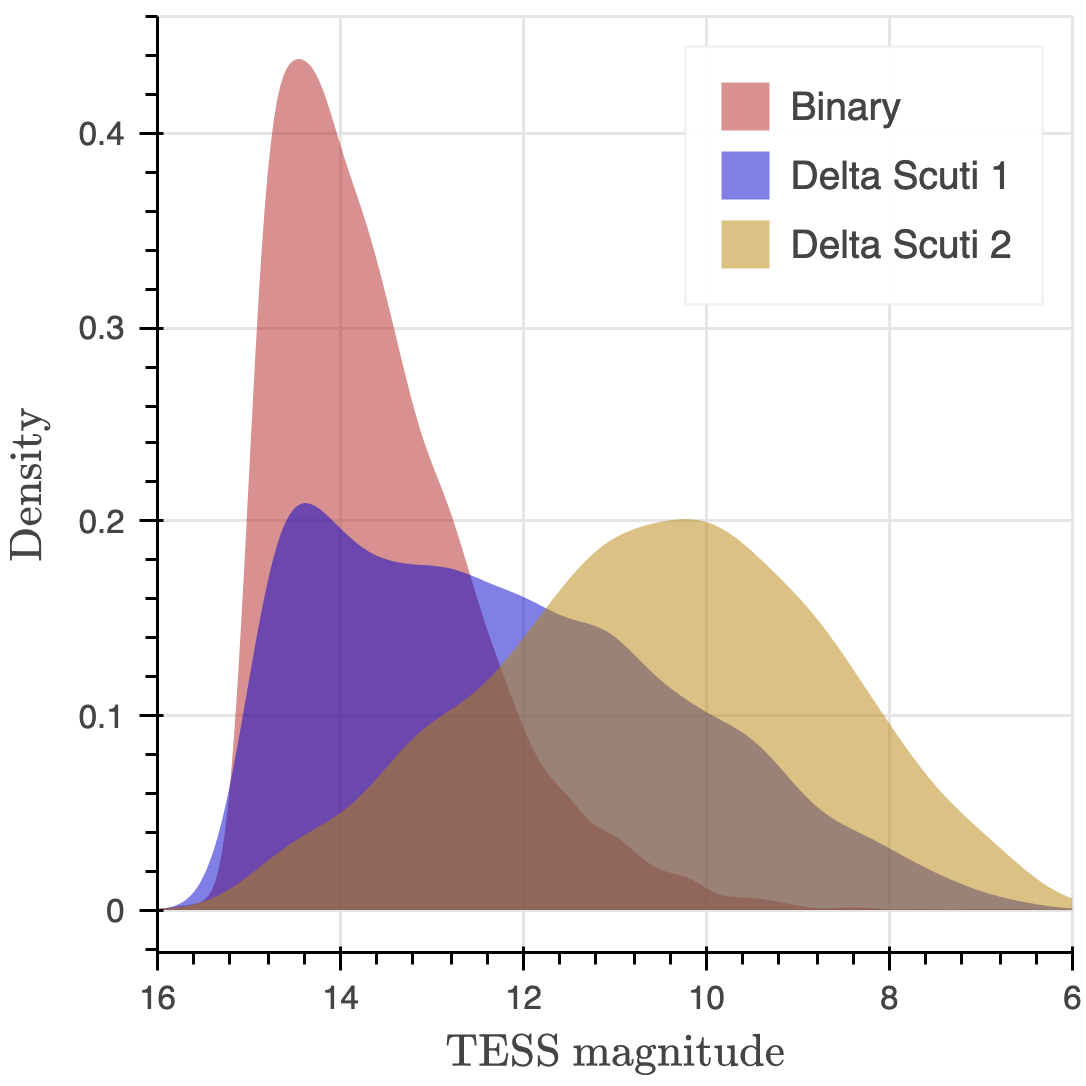}
    \caption{A kernel density estimation of the TESS magnitudes of the short-period variables. Magnitude values come from the TIC. Distributions are colored by the partitions described in \cref{subsec:partitioning_the_data}.}
    \label{fig:magnitude_kde_clusters_colored_figure}
\end{figure}

\begin{figure}
    \centering
    \begin{pycode}
        from results_resources.parallax_kde_clusters_colored_figure import generate_parallax_kde_clusters_colored_figure
        create_latex_figure_from_bokeh_figure(bokeh_figure=generate_parallax_kde_clusters_colored_figure(),
                                              latex_figure_path='results_resources/parallax_kde_clusters_colored_figure.png',
                                              latex_width='\columnwidth',
                                              latex_height='\columnwidth')
    \end{pycode}
    \includegraphics[width=\columnwidth]{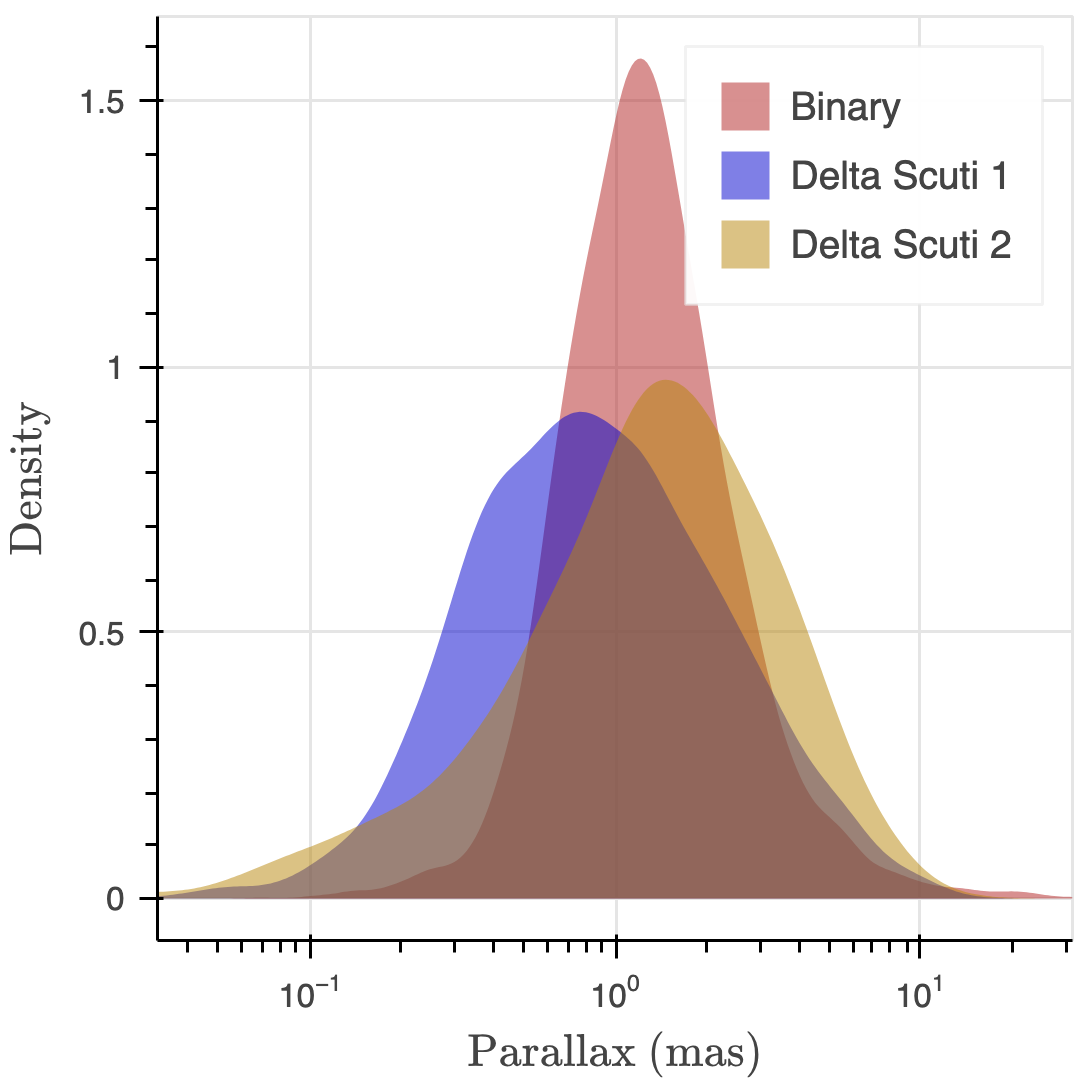}
    \caption{A kernel density estimation of the parallaxes of the short-period variables. Parallax values come from the TIC. Distributions are colored by the partitions described in \cref{subsec:partitioning_the_data}.}
    \label{fig:parallax_kde_clusters_colored_figure}
\end{figure}

After partitioning the data as described above, we then visualize the distributions of the raw observed values for each partitioned cluster of data. The observed distributions of the photometric color of each cluster is shown in \cref{fig:color_kde_clusters_colored_figure}. Similarly, \cref{fig:magnitude_kde_clusters_colored_figure} shows the distributions of the TESS magnitudes of the clusters and \cref{fig:parallax_kde_clusters_colored_figure} the distributions of the parallax.

 \subsection{Binaries analysis}\label{subsec:binaries_analysis}

The short-period \gls{MS} binary partition contains \numberOfTargetsInRedDwarfBinaryCluster{} targets, from which the overwhelming majority come from the population of binaries seen as the large grouping with longer periods in \cref{fig:absolute_v_magnitude_vs_log_period_clusters_colored_figure}. At the shorter period end of this population, there is a sharp cutoff at \approximately0.12 days. With binaries, the full orbital period is twice this value at \approximately0.24 days. This is consistent with a known sharp cut-off of the period distribution of red dwarf binaries at \approximately0.22 days~\citep{rucinski1992can,norton2011short}. This cut-off is also seen in \cref{fig:period_vs_radius_figure} which shows radius vs period. As expected, this lower period is only found with cooler, smaller binaries, as the hotter binaries are larger and cannot easily orbit as closely.

This can easily be understood from basic stellar structure scaling relationships. For F, G, K stars with $0.5\lesssim(M/M_\odot)\lesssim 2$, the radius scales roughly like $R=R_\odot (M/M_\odot)^{3/4}$. For an equal-mass binary $M_1=M_2$, Roche lobe overflow will occur when the binary separation is roughly $a\approx 2.6 R$, corresponding to a minimum orbital period of 
\begin{eqnarray}
    T_{\rm orb} &=& 2\pi\sqrt{\frac{a^3}{G(M_1+M_2)}} \nonumber \\
    &=& 2\pi\sqrt{8.8\frac{R_\odot^3}{G M_\odot}\left(\frac{M_1}{M_\odot}\right)^{5/4}} \nonumber \\
    &=& 0.34 \left(\frac{M_1}{M_\odot}\right)^{5/8} \mbox{ d}.
\end{eqnarray}
For a red dwarf binary with $M_1=M_2=0.5 M_\odot$, this corresponds to a binary period of 0.22d, just as we find empirically. Furthermore, if we use the mass-luminosity scaling for \gls{MS} stars $L\approx L_\odot (M/M_\odot)^3$, we are able to recover the relationship $L\sim T_{\rm orb}^{4.8}$, consistent with what we see in the lower track of Figure \ref{fig:absolute_v_magnitude_vs_log_period_clusters_colored_figure}.
\begin{figure}
    \centering
    \begin{pycode}
        from results_resources.fundamental_vs_first_overtone_figure import generate_fundamental_vs_first_overtone_figure
        create_latex_figure_from_bokeh_figure(bokeh_figure=generate_fundamental_vs_first_overtone_figure(),
                                              latex_figure_path='results_resources/fundamental_vs_first_overtone_figure.png',
                                              latex_width='\columnwidth',
                                              latex_height='\columnwidth')
    \end{pycode}
    \includegraphics[width=\columnwidth]{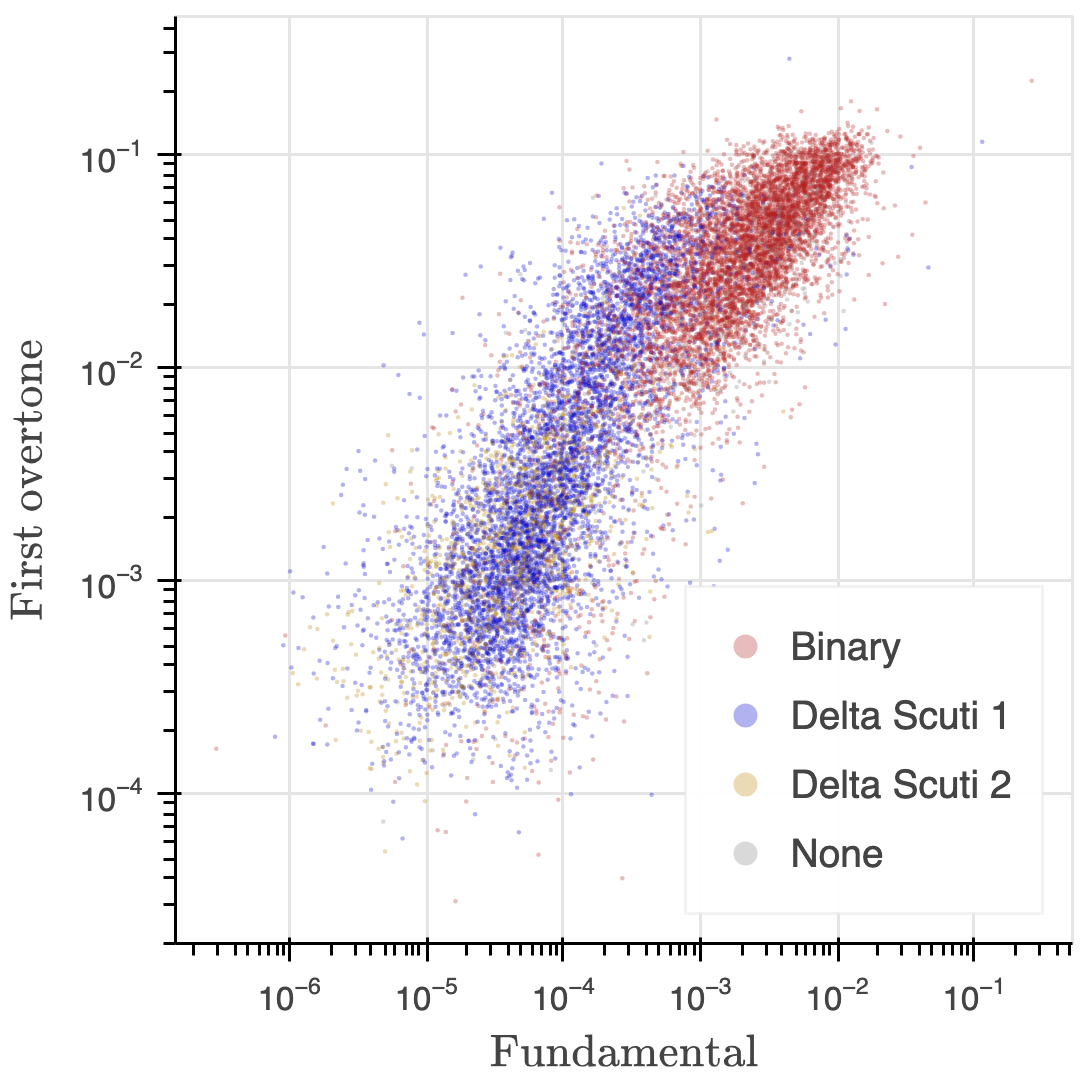}
    \caption{The powers of the fundamental and first overtone. Data points are colored by the partitions described in \cref{subsec:partitioning_the_data}.}
    \label{fig:fundamental_vs_first_overtone_figure}
\end{figure}

We folded the \glspl{LC} on a period $p_{fold} = 2 p_{estimated}$, where $p_{estimated}$ is the period estimated by our post-processing described in \cref{sec:post_neural_network_processing}. Then, on these folded \glspl{LC}, we applied a fast-Fourier transform, resulting in a fundamental mode with period $p_{fold}$ and the first overtone mode with period $p_{estimated}$. \cref{fig:fundamental_vs_first_overtone_figure} shows the powers of the fundamental mode and the first overtone mode. Notably, we expect an increased fundamental for the binaries, as the differences between the two stars will primarily affect this mode.

\begin{figure}
    \centering
    \begin{pycode}
        from results_resources.period_vs_relative_amplitude_figure import generate_period_vs_relative_amplitude_figure
        create_latex_figure_from_bokeh_figure(bokeh_figure=generate_period_vs_relative_amplitude_figure(),
                                              latex_figure_path='results_resources/period_vs_relative_amplitude_figure.png',
                                              latex_width='\columnwidth',
                                              latex_height='1.3\columnwidth')
    \end{pycode}
    \includegraphics[width=\columnwidth]{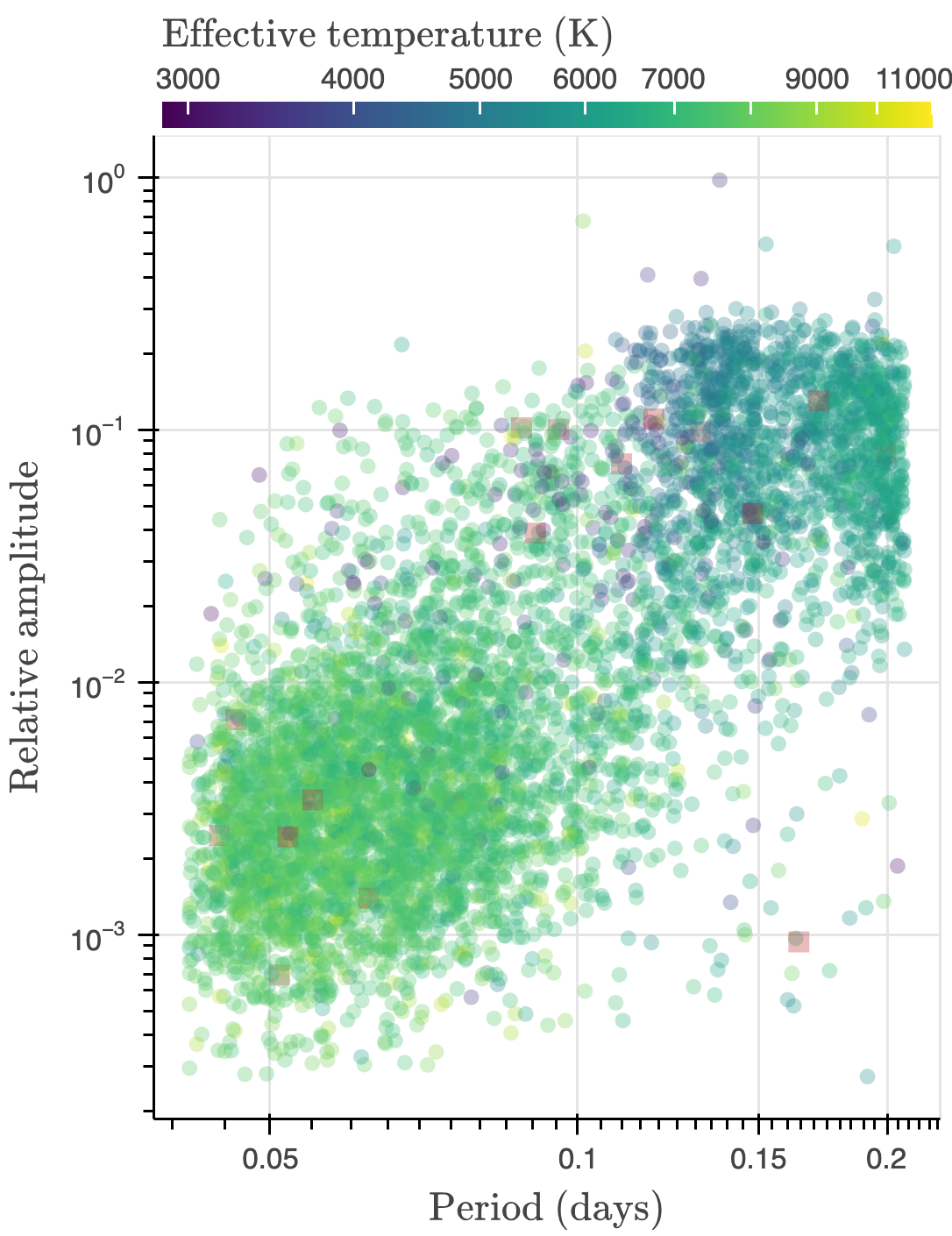}
    \caption{The relative amplitude vs the period of the short-period variables color by the variable effective temperature. Due to high effective temperature outliers, the upper limit of the effective temperature scale is restricted to \highTemperatureScaleLimit{}K and any targets above this effective temperature are shown as \highTemperatureColor{} squares.}
    \label{fig:period_vs_relative_amplitude_figure}
\end{figure}

In \cref{fig:period_vs_relative_amplitude_figure}, we see the relative amplitude of the periodic signal of the short-period variables, again split into two primary populations. The top right population is the binary population, with amplitudes that mostly range from \approximately0.02 to ~\approximately0.3. These estimates only apply to targets with a known contamination ratio from the \gls{TIC}. As also evident in \cref{fig:fundamental_vs_first_overtone_figure}, the binary cluster targets have, on average, a much larger relative amplitude than the \gls{DS} cluster targets.

 \subsection{Delta Scuti analysis}\label{subsec:delta_scuti_analysis}

\begin{figure}
    \centering
    \begin{pycode}
        from results_resources.period_vs_radius_clusters_colored_figure import generate_period_vs_radius_clusters_colored_figure
        create_latex_figure_from_bokeh_figure(bokeh_figure=generate_period_vs_radius_clusters_colored_figure(),
                                              latex_figure_path='results_resources/period_vs_radius_clusters_colored_figure.png',
                                              latex_width='\columnwidth',
                                              latex_height='1.3\columnwidth')
    \end{pycode}
    \includegraphics[width=\columnwidth]{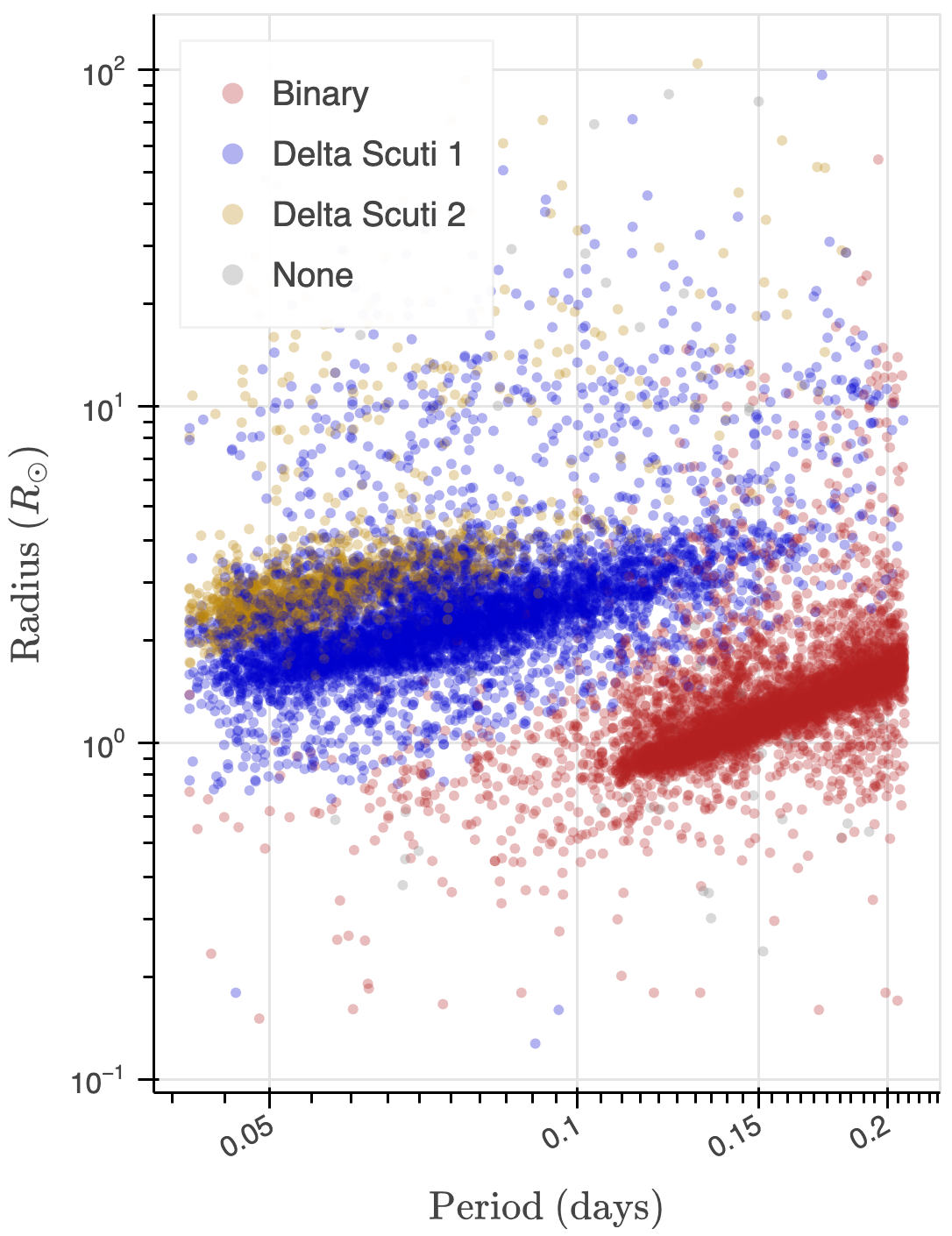}
    \caption{The radius vs the period of the short-period variables. Data points are colored by the partitions described in \cref{subsec:partitioning_the_data}.}
    \label{fig:period_vs_radius_clusters_colored_figure}
\end{figure}

Similar to \cref{fig:period_vs_radius_figure}, \cref{fig:period_vs_radius_clusters_colored_figure} shows the targets in radius vs period space. \cref{fig:period_vs_radius_figure} is useful for seeing the gap between the two \gls{DS} populations, while \cref{fig:period_vs_radius_clusters_colored_figure} shows the partition coloring after the partitioning has been made.

\begin{figure}
    \centering
    \begin{pycode}
        from results_resources.temperature_vs_luminosity_clusters_colored_figure import generate_temperature_vs_luminosity_clusters_colored_figure
        create_latex_figure_from_bokeh_figure(bokeh_figure=generate_temperature_vs_luminosity_clusters_colored_figure(),
                                              latex_figure_path='results_resources/temperature_vs_luminosity_clusters_colored_figure.png',
                                              latex_width='\columnwidth',
                                              latex_height='\columnwidth')
    \end{pycode}
    \includegraphics[width=\columnwidth]{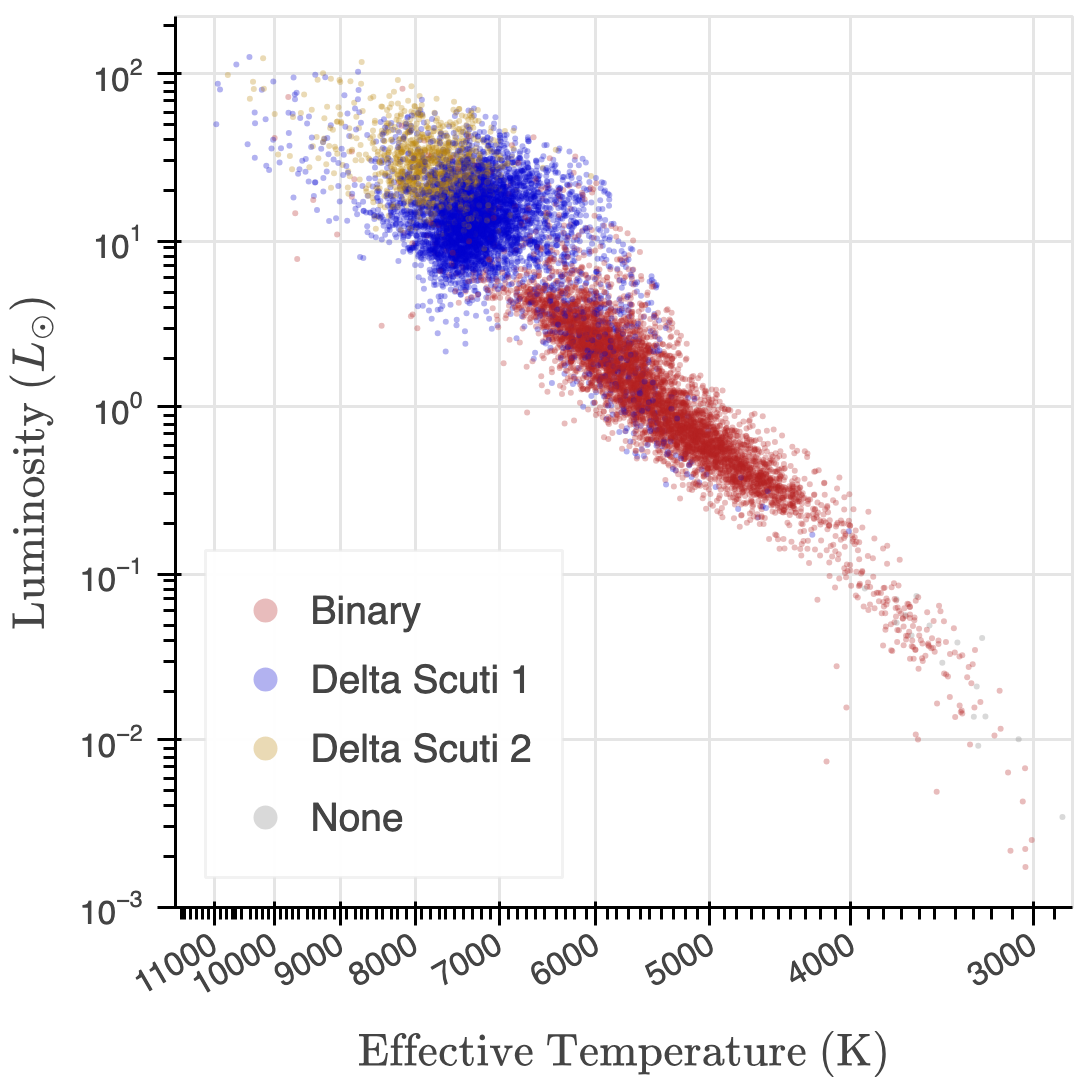}
    \caption{The effective temperature vs luminosity of the short-period variables. Distributions are colored by the partitions described in \cref{subsec:partitioning_the_data}.}
    \label{fig:temperature_vs_luminosity_clusters_colored_figure}
\end{figure}

\cref{fig:temperature_vs_luminosity_clusters_colored_figure} shows the effective temperature vs luminosity of the short-period variables. Notably, the \gls{DS} targets are higher along the \gls{MS} branch than the short-period \gls{MS} binaries, which is expected, and the partitioning chosen divides \gls{DS} from the \gls{MS} binaries fairly neatly.

Other works, including \citet{ziaali2019period,barac2022revisiting}, have shown a gap between the two \gls{DS} populations in log period space with a notable sparse region between the two populations, where the second ridge of \gls{DS} stars is known to have period of 0.5 that of primary ridge~\citep{ziaali2019period,barac2022revisiting}. The reasoning for the existence of the second ridge is not known. The gap is usually shown via the distribution of the distances in log period space to the partition line between the two populations, or an equivalent metric. The gap between the two ridges is not as clearly defined in our results as in \citet{ziaali2019period,barac2022revisiting}, with our distribution resulting in a short plateau in the distribution rather than a sparse region. However, we note that we have not performed a fitting to maximize this gap. Due to non-\gls{DS} targets certainly being within the presented clusterings and the uncertain detection efficiency of the neural network on each of the clusters, we don't expect such a fitting to produce an accurate representation of the two \gls{DS} populations. We provide these partitions only as a simple qualitative tool for comprehension. Based on this partitioning, our \gls{DS} partitions contain \numberOfTargetsInBothDeltaScutiClusters{} identified targets, with \numberOfTargetsInDeltaScutiPrimaryRidgeCluster{} in the primary ridge partition and \numberOfTargetsInDeltaScutiSecondRidgeCluster{} in the second ridge partition.
 \subsection{Hot targets analysis}\label{subsec:hot_targets_analysis}

Our \gls{NN} detected nine hot periodic objects with effective temperatures ranging from 28,000 to 38,000 Kelvins. These appear to be \gls{sdB} stars, which are extreme horizontal branch stars with very high surface gravity and temperature~\citep{Baran2023b,Baran2023a,schaffenroth2022hot}.

There are three channels to formation of \gls{sdB} stars involving close binary stellar pairs: common envelope, \gls{RLOF}, and \gls{WD} mergers~\citep{han2002origin}. The common envelope channel produces short-period \gls{WD}+\gls{sdB} or \gls{sdB}+\gls{MS} binaries with periods ranging from 0.1 to 10 days. \gls{TIC} IDs 333419799, 270491267, 396004353, 193092806, and 458785169 fall into this broad range. While the \gls{RLOF} channel produces stars with periods exceeding 400 days and are thus not addressed by this work, the \gls{WD} merger channel may produce single \gls{sdB} stars that exhibit asteroseismic vibrational modes with periods of less than 0.1 days. Four hot periodic objects detected by our \gls{ML} pipeline fall into this last category: \gls{TIC} IDs 173295499, 409644971, 99641129, and 367779738, with all but the final one having periods less than but very near 0.1 days. To be the best of our knowlege, \gls{TIC} ID 367779738 has not previously been identified as a subdwarf in a close binary, but the remainder of these targets have previously been identified as subdwarfs in close binaries~\citep{Baran2023b,Baran2023a,schaffenroth2022hot}. All of these targets are identified spectroscopically as hot subdwarf stars by the Large Sky Area Multi-Object Fibre Spectroscopic Telescope (LAMOST) data release 8~\citep{lei2023}.
 \subsection{Miscellaneous analysis}\label{subsec:miscellaneous}

\begin{figure}
    \centering
    \begin{pycode}
        from results_resources.period_vs_mass_figure import generate_period_vs_mass_figure
        create_latex_figure_from_bokeh_figure(bokeh_figure=generate_period_vs_mass_figure(),
                                              latex_figure_path='results_resources/period_vs_mass_figure.png',
                                              latex_width='\columnwidth',
                                              latex_height='1.3\columnwidth')
    \end{pycode}
    \includegraphics[width=\columnwidth]{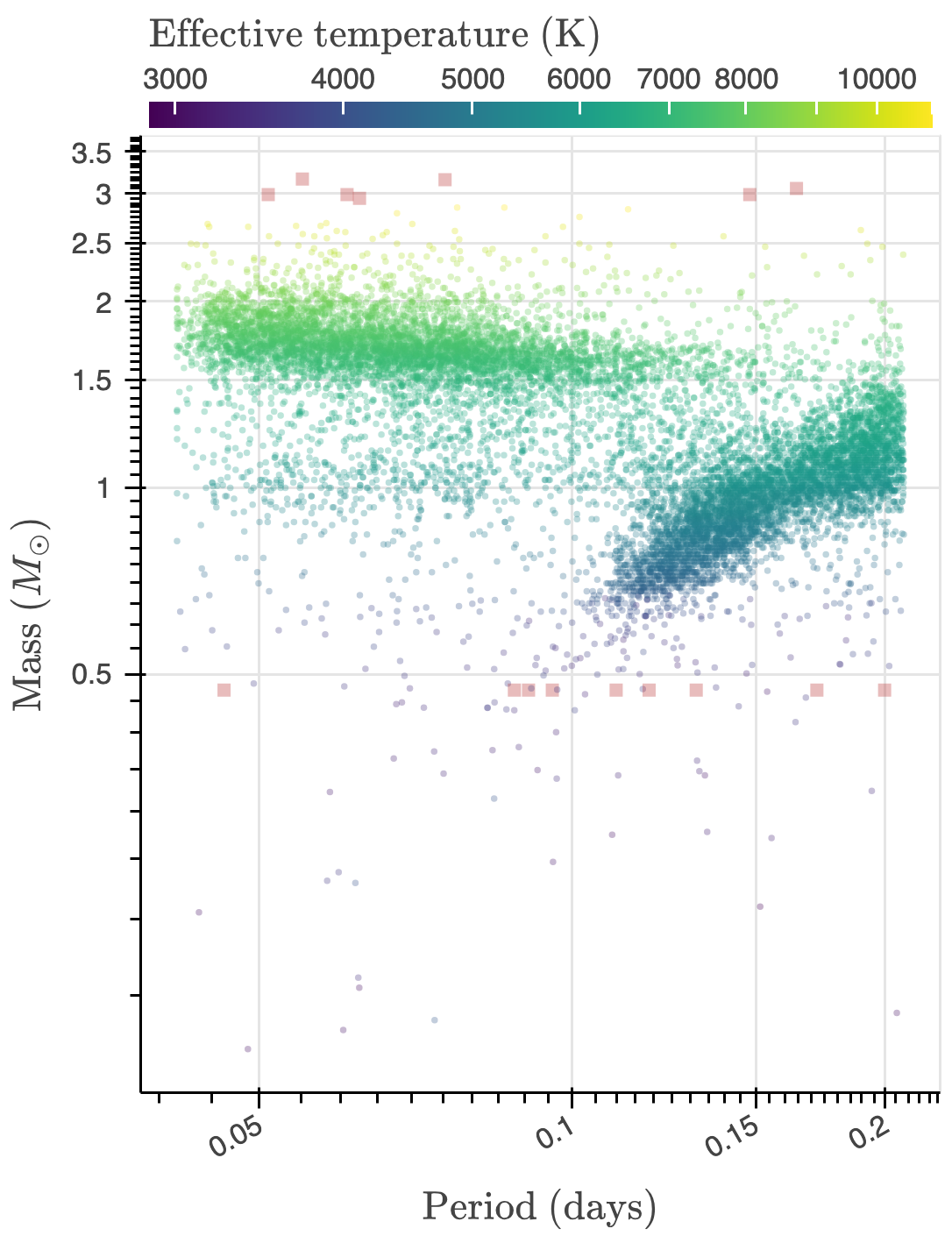}
    \caption{The mass vs the period of the short-period variables. Due to high effective temperature outliers, the upper limit of the effective temperature scale is restricted to \highTemperatureScaleLimit{}K and any targets above this effective temperature are shown as \highTemperatureColor{} squares.}
    \label{fig:period_vs_mass_figure}
\end{figure}

\cref{fig:period_vs_mass_figure} shows the mass vs period of the short-period variables. There is a sizable population of targets below the major \gls{DS} population. From our three partitions, most of this population comes from the \gls{DS} primary ridge population. However, this population may be a separate category of objects, or may simply be a group of targets whose mass is poorly estimated by the approach used by the \gls{TIC}.

      \section{Conclusion}\label{sec:conclusion}

In this work, we presented a \gls{NN} pipeline to identify short-period variable targets in \gls{TESS} \gls{FFI} data. This \gls{NN} pipeline performs inference on a \gls{TESS} 30-minute cadence light curve in \approximately5ms on a single GPU. We also presented a collection of \numberOfShortPeriodVariables{} short-period variables, and provided a cursory analysis of the various populations of these targets. Within these populations, we examine two prominent populations consisting primary of short-period \gls{MS} binaries and the \gls{DS} stars. \numberOfTargetsInRedDwarfBinaryCluster{} targets were identified in the \gls{MS} binary partition and \numberOfTargetsInBothDeltaScutiClusters{} were identified in the \gls{DS} partitions. Though not all, we expect the overwhelming majority of targets in each partition to be of the class that is used to name each partition. We provide this collection of targets in a supplemental file for more detailed analysis by specialists in the field for each of the populations.

For future work, our open-source \gls{NN} framework allows for this work to be relatively easily extended. Notably, this work uses the \gls{TESS} primary mission \gls{FFI} data which had a 30-minute cadence. Later \gls{TESS} \gls{FFI} data has a higher cadence, which would enable searches for shorter period targets and more accurate post-processing. A simple change to the \gls{NN} (\eg, adjusting the stride lengths) would prepare it for training on this shorter cadence dataset. This, combined with the relatively fast inference speed of our network (\approximately{}5ms per \gls{LC} on a single GPU), allows our method to be quickly deployed on new datasets.     \section{Acknowledgments}

This paper includes data collected by the {\em TESS} mission, which are publicly available from the Mikulski Archive for Space Telescopes (MAST). Funding for the {\em TESS} mission is provided by NASA's Science Mission directorate.

Resources supporting this work were provided by the NASA High-End Computing (HEC) Program through the NASA Center for Climate Simulation (NCCS) at Goddard Space Flight Center.

This work has made use of data from the European Space Agency (ESA) mission \textit{Gaia} (\url{https://www.cosmos.esa.int/gaia}), processed by the \textit{Gaia} Data Processing and Analysis Consortium (DPAC, \url{https://www.cosmos.esa.int/web/gaia/dpac/consortium}).

The material is based upon work supported by NASA under award number 80GSFC21M0002.

\facilities{
    \textit{Gaia},
    MAST,
    NCCS,
    \textit{TESS}
}

\software{
    \texttt{Astropy}~\citep{astropy2013,astropy2018},
    \texttt{Bokeh}~\citep{bokeh},
    \texttt{Eleanor}~\citep{feinstein2019eleanor},
    \texttt{Keras}~\citep{keras},
    \texttt{Lightkurve}~\citep{lightkurve},
    \texttt{Matplotlib}~\citep{matplotlib},
    \texttt{NumPy}~\citep{numpy},
    \texttt{Pandas}~\citep{pandas},
    \texttt{pytest}~\citep{pytestx.y},
    \texttt{Python}~\citep{python38},
    \texttt{Tensorflow}~\citep{tensorflow}
}
     \bibliography{bibliography}
\bibliographystyle{aasjournal}
     \FloatBarrier
{
    \begin{pycode}
        from results_resources.random_sample_table import generate_random_sample_table_file
        generate_random_sample_table_file()
    \end{pycode}
    \begin{turnpage}
    \centering
    \begin{table}
\centering
\caption{A random sample of the short-period variables table. Due to page space constraints, some columns are excluded here which are include in the supplemental file. For the partitions, 0 corresponds to the binary partition, 1 corresponds to the $\delta$ Sct primary partition, and 2 corresponds to the $\delta$ Sct secondary partition.}
\label{tab:short_period_variables_table}
\begin{tabular}{rrrrrrrrrrr}
\toprule
$\begin{tabular}{@{}c@{}}TIC ID\end{tabular}$ & $\begin{tabular}{@{}c@{}}Period\\(days)\end{tabular}$ & $\begin{tabular}{@{}c@{}}TESS\\magnitude\end{tabular}$ & $\begin{tabular}{@{}c@{}}Color\\(B-V)\end{tabular}$ & $\begin{tabular}{@{}c@{}}Parallax\\(mas)\end{tabular}$ & $\begin{tabular}{@{}c@{}}Absolute V\\magnitude\end{tabular}$ & $\begin{tabular}{@{}c@{}}Effective\\temperature\\(K)\end{tabular}$ & $\begin{tabular}{@{}c@{}}Radius\\($R_\odot$)\end{tabular}$ & $\begin{tabular}{@{}c@{}}Mass\\($M_\odot$)\end{tabular}$ & $\begin{tabular}{@{}c@{}}Relative\\amplitude\\(contamination\\corrected)\end{tabular}$ & $\begin{tabular}{@{}c@{}}Partition\end{tabular}$ \\
\midrule
27399756 & 0.191597 & 14.3084 & 0.432 & 0.422676 & 3.279721 & 6206 & 1.99035 & 1.190 & N/A & 0 \\
57104837 & 0.118362 & 12.5353 & 0.944 & 3.488980 & 6.298421 & 4688 & 0.95951 & 0.748 & 0.010296 & 0 \\
63395663 & 0.048016 & 7.7804 & -0.026 & 4.348280 & 1.185597 & 7904 & 2.87852 & 1.892 & 0.004903 & 2 \\
84079248 & 0.133662 & 14.3848 & 0.961 & 1.592000 & 6.607015 & 4844 & 0.98444 & 0.790 & N/A & 0 \\
85397534 & 0.041669 & 13.3796 & 0.192 & 0.428423 & 1.991264 & 7818 & 2.16868 & 1.860 & N/A & 1 \\
129011402 & 0.059596 & 10.0443 & 0.167 & 2.768110 & 2.603268 & 7409 & 1.77860 & 1.690 & 0.001290 & 1 \\
150395859 & 0.129224 & 14.2772 & 1.170 & 1.349820 & 5.557480 & 4987 & 0.96169 & 0.830 & N/A & 0 \\
154089766 & 0.159397 & 13.4165 & 0.473 & 1.296250 & 4.624436 & 5747 & 1.13321 & 1.030 & N/A & 0 \\
158321712 & 0.136393 & 9.1802 & 0.332 & 2.426560 & 1.546268 & 6954 & 3.17292 & 1.520 & 0.015252 & 1 \\
164143903 & 0.178556 & 10.5526 & 0.444 & 2.956590 & 3.483750 & 6012 & 1.78928 & 1.111 & 0.074664 & 0 \\
198457457 & 0.201747 & 9.4010 & 0.493 & 4.555980 & 3.115612 & 6565 & 1.67415 & 1.362 & 0.092000 & 0 \\
202129305 & 0.066931 & 13.8895 & 0.657 & 0.414300 & 2.553055 & 6711 & 2.44164 & 1.430 & N/A & 1 \\
229968259 & 0.065755 & 11.5932 & 0.325 & 1.334230 & 2.338049 & 7262 & 1.80141 & 1.640 & 0.012970 & 1 \\
240936297 & 0.055294 & 7.6339 & 0.169 & 7.534510 & 2.183399 & 8086 & 1.71841 & 1.964 & 0.000989 & 1 \\
250190926 & 0.185086 & 13.5371 & 0.407 & 0.538793 & 2.693453 & 6502 & 2.13631 & 1.330 & N/A & 0 \\
259895398 & 0.085386 & 14.7899 & 0.530 & N/A & N/A & 4958 & N/A & N/A & N/A & N/A \\
279161696 & 0.145065 & 11.6053 & 0.641 & 3.533400 & 5.071595 & 5318 & 1.14720 & 0.920 & 0.050170 & 0 \\
287458386 & 0.055065 & 14.2555 & 0.451 & 0.482627 & 3.226748 & 7297 & 1.59176 & 1.650 & N/A & 1 \\
289563235 & 0.053302 & 11.9361 & 0.792 & 1.351010 & 3.173538 & 7029 & 2.01295 & 1.550 & 0.002363 & 1 \\
302280281 & 0.155104 & 13.9746 & 0.461 & 1.036030 & 4.713142 & 5741 & 1.16163 & 1.030 & N/A & 0 \\
306738858 & 0.063785 & 9.6572 & 0.299 & 2.973500 & 2.253192 & 7781 & 1.90799 & 1.840 & 0.001603 & 1 \\
329248237 & 0.127547 & 14.6418 & 0.467 & 1.352740 & 6.322457 & 4828 & 0.88212 & 0.780 & N/A & 0 \\
330331931 & 0.076383 & 12.7128 & 0.706 & 0.977406 & 3.425832 & 6416 & 2.26155 & 1.290 & 0.059482 & 1 \\
348363727 & 0.131109 & 14.7367 & 1.345 & 0.012004 & 1.378569 & 4587 & 9.88582 & N/A & N/A & 1 \\
367419535 & 0.200030 & 13.3578 & 1.152 & 0.256559 & 1.819218 & 4646 & 8.94662 & N/A & N/A & 0 \\
368313899 & 0.109381 & 13.7293 & 0.552 & 0.816118 & 4.297064 & 6033 & 1.87990 & 1.120 & N/A & 0 \\
394171494 & 0.084973 & 13.3296 & -0.650 & 2.136530 & 5.725674 & 5334 & 0.83553 & 0.920 & N/A & 0 \\
453184196 & 0.163674 & 13.5680 & 1.017 & 0.406524 & 2.616653 & 6037 & 4.04076 & 1.120 & N/A & 0 \\
453285891 & 0.147838 & 12.2257 & 0.716 & 3.463850 & 6.298779 & 4955 & 1.13484 & 0.820 & 0.124630 & 0 \\
468663462 & 0.193995 & 14.4012 & 0.742 & N/A & 4.902140 & 5109 & N/A & N/A & N/A & 0 \\
\bottomrule
\end{tabular}
\end{table}
     \end{turnpage}
}
 \end{document}